\documentclass[
  letterpaper,
	final,
	aps,
	twocolumn,
	]{revtex4-1}
\let\Twocolumn

\newif\ifTwocolumn
\ifx\Twocolumn\undefined
  \Twocolumnfalse
\else
  \Twocolumntrue
\fi
\ifTwocolumn
\fi
\usepackage{graphicx}
\usepackage{amsmath,amssymb,amsfonts}
\usepackage{gensymb}
\usepackage{bm}
\usepackage{upgreek}
\usepackage{wasysym}
\usepackage{bibentry}
\usepackage[T1]{fontenc}
\usepackage{textcomp}
\usepackage{mathptmx}
\usepackage[scaled=0.92]{helvet}
\usepackage{courier}
\usepackage{color}
\newcommand{\eref}[1]{Eq.~\eqref{#1}}
\newcommand{\fref}[1]{Fig.~\ref{#1}}

\newcommand{\upd}{{\ensuremath{\textrm{d}}}}
\renewcommand*{\vec}[1]{\mathbf{#1}}

\newcommand{\Mm}{\ensuremath{(-,-)}}
\newcommand{\Pp}{\ensuremath{(+,+)}}
\newcommand{\Pm}{\ensuremath{(+,-)}}

\newcommand{\Ab}{\ensuremath{(a,b)}}
\newcommand{\Asb}{\ensuremath{ {(a_<,b)}}}
\newcommand{\Alb}{\ensuremath{ {(a_>,b)}}}
\newcommand{\step}{\ensuremath{\textrm{s}}}

\newcommand{\Fab}{\ensuremath{F_{\Ab}}}

\newcommand{\Fpm}{\ensuremath{F_{\Pm}}}

\newcommand{\Fstep}{\ensuremath{F_{\step}}}

\newcommand{\kpm}{\ensuremath{k_{\Pm}}}
\newcommand{\kmm}{\ensuremath{k_{\Mm}}}

\newcommand{\kab}{\ensuremath{k_{\Ab}}}

\newcommand{\Kpm}{\ensuremath{K_{\Pm}}}
\newcommand{\Kmm}{\ensuremath{K_{\Mm}}}

\newcommand{\Kab}{\ensuremath{K_{\Ab}}}
\newcommand{\Kasb}{\ensuremath{K_{\Asb}}}
\newcommand{\Kalb}{\ensuremath{K_{\Alb}}}

\newcommand{\Kstep}{\ensuremath{K_{\step}}}

\newcommand{\Phiab}{\ensuremath{\Phi_{\Ab}}}
\newcommand{\Phistep}{\ensuremath{\Phi_{\step}}}

\newcommand{\varthetaab}{\ensuremath{\vartheta_{\Ab}}}
\newcommand{\varthetaasb}{\ensuremath{\vartheta_{\Asb}}}
\newcommand{\varthetaalb}{\ensuremath{\vartheta_{\Alb}}}
\newcommand{\varthetastep}{\ensuremath{\vartheta_{\step}}}

\newcommand{\Dpp}{\ensuremath{\Delta_{\Pp}}}
\newcommand{\Dmm}{\ensuremath{\Delta_{\Mm}}}
\newcommand{\Dpm}{\ensuremath{\Delta_{\Pm}}}

\newcommand{\Dab}{\ensuremath{\Delta_{\Ab}}}

\newcommand{\ddef}{\ensuremath \mathrel{\mathop:}=}

\newcommand{\Len}{\ensuremath{\mathcal{L}}}
\newcommand{\lat}{\ensuremath{\ell}}
\DeclareMathOperator\sgn{sign}

\DeclareMathOperator\erf{erf}
\ifTwocolumn
  
\else
  
\fi

\usepackage{setspace}
\ifTwocolumn
\newcommand{\clevercaption}[1]{\caption{\setstretch{1}#1}}
\else
\newcommand{\clevercaption}[1]{\caption{\setstretch{1.7}#1}}
\fi

\begin{document}
\setstretch{1.05}
\title{Alignment of cylindrical colloids near chemically\\ patterned substrates induced by critical Casimir torques}
\author{M.~Labb{\'e}-Laurent}
\email{laurent@is.mpg.de}
\author{M.~Tr{\"o}ndle}
\email{troendle@is.mpg.de}
\author{L.~Harnau}
\email{harnau@is.mpg.de}
\author{S.~Dietrich}
\email{dietrich@is.mpg.de}
\affiliation{
	Max-Planck-Institut f\"ur Intelligente Systeme,  
	Heisenbergstr.\ 3, D-70569 Stuttgart, Germany}
\affiliation{
	IV. Institut f\"ur Theoretische Physik, 
	Universit\"at Stuttgart, 
	Pfaffenwaldring 57, 
	D-70569 Stuttgart, Germany
}
\date{\today}
%
\begin{abstract}\setstretch{1.2}
  Recent experiments have demonstrated a fluctuation-induced lateral trapping of spherical colloidal particles immersed
  in a binary liquid mixture near its critical demixing point and exposed to chemically
  patterned substrates.
  Inspired by these experiments, we study this kind of effective interaction, known as the
  critical Casimir effect, for elongated colloids of cylindrical shape.
  This adds orientational degrees of freedom.
  When the colloidal particles are close to a chemically structured substrate, a critical
  Casimir torque acting on the colloids emerges.
  We calculate this torque on the basis of the Derjaguin approximation.
  The range of validity of the latter is assessed
  via mean-field theory.
  This assessment shows that the Derjaguin approximation is reliable in experimentally relevant 
  regimes, so that we extend it to Janus particles endowed with opposing adsorption preferences.
  Our analysis indicates that critical Casimir interactions are capable of achieving well-defined, reversible alignments both of chemically homogeneous and of Janus cylinders.
\end{abstract}
\maketitle

\section{Introduction \label{sec:introduction}}
%
\begin{figure*}[t!]
    \centering
    \ifTwocolumn
    \includegraphics[width=12cm]{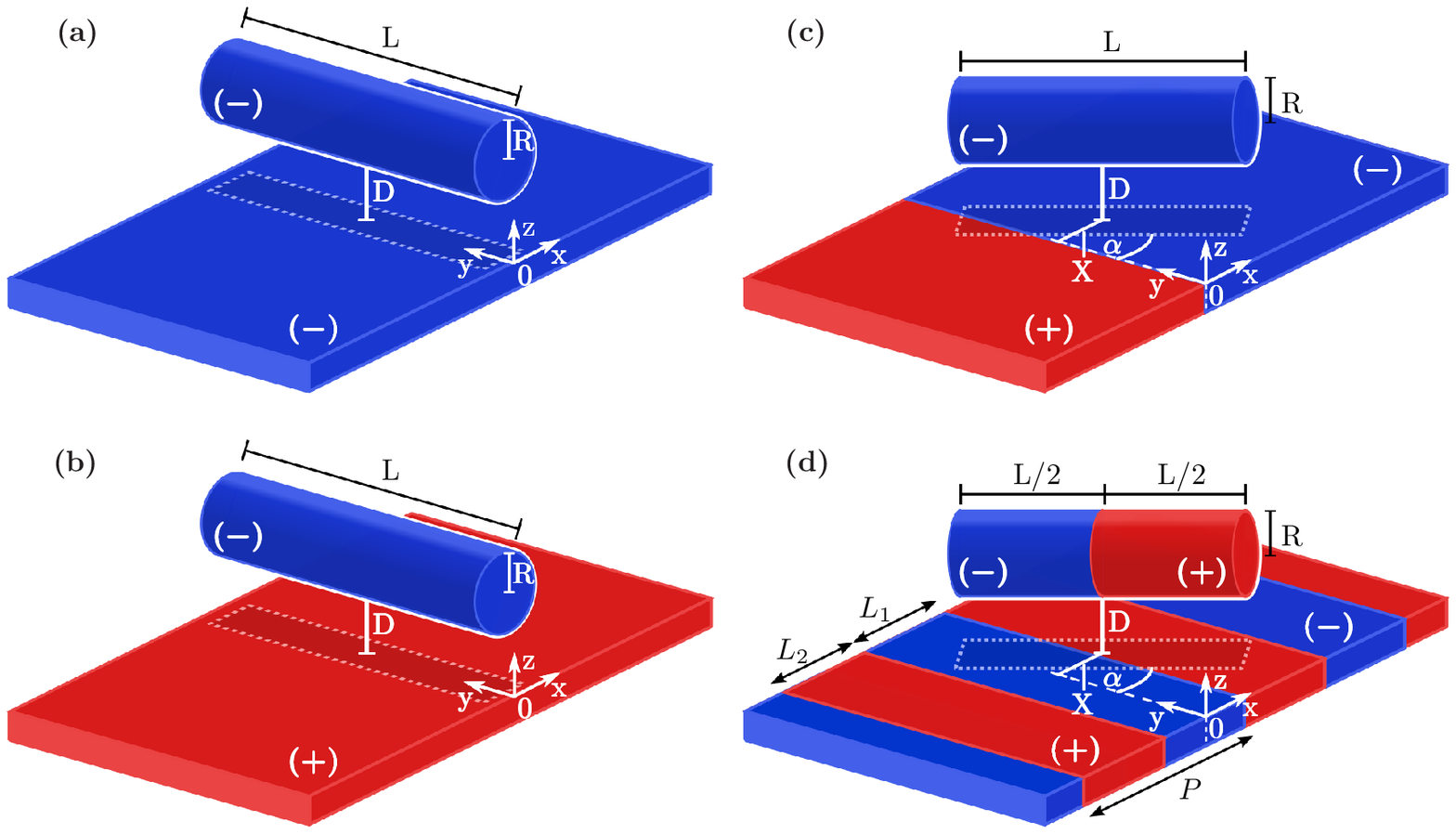}
    \else
    \includegraphics[width=\textwidth]{fig_01}
    \fi
    \clevercaption{Illustration of the geometries and the boundary conditions (BCs) considered in our study: 
    (a) cylinder of radius $R$ and length $L$ oriented parallel to a substrate at a surface-to-surface distance $D$, both surfaces exhibiting the same homogeneous
    adsorption preference for the two species of a binary liquid mixture corresponding to $(-)$ BC,
    (b) cylinder close to a homogeneous substrate which exhibits the opposite adsorption preference corresponding to  $(+)$ BC,
    (c) cylinder close to a chemical step at which the adsorption preference of the substrate changes discontinuously along 
    the lateral direction $x$ from $(+)$ BC at $x<0$ to $(-)$ BC at $x\ge0$ so that the cylinder axis is rotated by an angle $\alpha\in[0,\pi/2]$ with respect to the chemical step at $x=0$, and
    (d) Janus cylinder close to a periodically striped substrate. The Janus cylinder exhibits opposing $(+)$ and $(-)$ BC at its two halves, and the substrate consists of stripes of width $L_1$ with
    $(-)$ BC neighboring stripes of width $L_2$ with $(+)$ BCs such that the periodicity is $P=L_1+L_2$. Moreover, the cylinder is rotated by an angle $\alpha$ with respect to the chemical steps of the stripes.
    In all cases the vertical projection of the cylinder onto the substrate surface forms a $2R\times L$ rectangle (dashed lines). 
    The center of the colloid is located at (a and b) $(x,y,z)=(0,L/2,D+R)$ and (c and d) $(x,y,z)=(X,L/2,D+R)$.
    }
    \label{fig:cyl_substrate}
\end{figure*}

The first direct measurement of critical Casimir forces acting on colloidal particles \cite{hertlein:2008} 
has recently demonstrated that in soft matter thermal fluctuations can induce well-directed effective interactions. 
Since its prediction \cite{fisher:1978}, the critical Casimir effect has attracted numerous theoretical
investigations (see, e.g., Refs.~\cite{krech:book} and \cite{brankov:book} and references therein) revealing both an intriguing
instance of solvent-mediated effective interactions and a classical analogue of the celebrated Casimir effect
in quantum electrodynamics \cite{casimir:1948,kardar:1999,capasso:2007,gambassi:2009conf}.

The critical Casimir effect arises due to the confinement of local density or concentration fluctuations
of a liquid by surfaces provided, e.g., by immersed colloids or substrates.
Whereas in general the spatial extent of fluctuations, given by the bulk correlation length $\xi$, is on the molecular scale,
it diverges upon approaching a critical point in the phase diagram of the fluid and attains mesoscopic values.
Thus, along such a thermodynamic path the effects of the individual confinements start to interfere and an effective interaction acting on the 
immersed objects sets in. Its amplitude is proportional to $k_BT$, where $k_B$ is Boltzmann's constant and $T$ is temperature.

Due to the algebraic divergence ${\xi(T\to T_c)\propto|T-T_c|^{-\nu}}$ of the correlation length, where
$T_c$ is the critical temperature and $\nu$ is a standard bulk critical exponent, the strength of 
the critical Casimir force responds to minute temperature changes.
These fluctuation-induced interactions can be described in terms of universal scaling functions determined by the bulk and 
surface universality classes of the system \cite{binder:1983,diehl:1986,diehl:1997}. 
Simple fluids and binary liquid mixtures belong to the \emph{Ising} universality
class.
At the confining surfaces these fluids are generically exposed to boundary conditions (BCs) belonging to the so-called 
\emph{normal} surface universality class.
Depending on the adsorption preference for one of the two species of a binary liquid mixture, the normal 
universality class allows for two possible boundary conditions denoted as $(+)$ or $(-)$ BC.
A colloidal particle is attracted by a nearby substrate if their surfaces share the same BCs, whereas
it is repelled if the mutual adsorption preferences are opposite.
Critical Casimir forces, therefore, provide a tool to design the strength and the direction of 
mesoscopic colloidal interactions via suitable surface treatment and temperature control \cite{gambassi:2009news}.

Experimentally, critical Casimir forces have been first measured indirectly by studying the thickness
of thin wetting films upon passing thermodynamically a critical end point in the phase diagram of the fluid
\cite{nightingale:1985,krech:9192all}. 
For classical binary liquid mixtures \cite{fukuto:2005,rafai:2007} as well as for
quantum binary liquids $^3$He~/~$^4$He \cite{garcia:2002,ueno:2003} and liquid $^4$He \cite{garcia:1999,ganshin:2006}
such experiments have provided evidence of the critical Casimir effect in agreement
with theoretical results obtained via {Monte} {Carlo} (MC) simulations \cite{hucht:2007,vasilyev:2007,vasilyev:2009,hasenbusch:2009b,hasenbusch:2010,hasenbusch:2010a}.

The critical Casimir effect in colloidal suspensions has attracted numerous 
theoretical investigations \cite{Burkhardt:1995,Eisenriegler:1995,hanke:1998,Schlesener:2003,Eisenriegler:2004}
due to the wide use of colloids in applications and due to their property as a paradigmatic soft matter model.
Using colloids, critical Casimir forces have been measured directly by monitoring
the position of a single spherical colloidal particle immersed in a binary liquid mixture of water 
and 2,6-lutidine and close to chemically homogeneous substrates
\cite{hertlein:2008,gambassi:2009,nellen:2009}.
The experimental data agree very well with theoretical predictions \cite{hertlein:2008,gambassi:2009,pousaneh:2012},
which are obtained via the Derjaguin approximation (DA) \cite{derjaguin:1934} and make use of the aforementioned
MC simulation results for the film geometry. 
A full MC simulation study for the sphere-wall geometry is numerically challenging and has been 
performed only recently \cite{hasenbusch:2013}, still limited to small ratios of the radius of the sphere
and the surface-to-surface distance between the wall and the sphere.

The combination of attractive and repulsive critical Casimir forces, as obtained by means of \emph{chemically patterning} substrates
with stripes of alternating adsorption preference, generates \emph{lateral} critical Casimir forces acting
on colloids, which have recently been realized experimentally \cite{soyka:2008,troendle:2011}. 
Laterally confining potentials, which can reversibly trap colloids, have been measured and agree very well 
with the corresponding theoretical predictions \cite{troendle:2009,troendle:2011}.
Critical Casimir forces between homogeneous and chemically striped surfaces have been studied theoretically
within mean-field theory (MFT) \cite{sprenger:2006,parisen:2013} and by using MC 
simulations \cite{parisen:2010,parisen:2010a,parisen:2013} for the film geometry.
The critical Casimir interaction between a spherical colloid and a chemically patterned
surface has been studied by some of the present authors \cite{troendle:2010}.
In Ref.~\cite{troendle:2010} normal and lateral critical Casimir forces have been calculated 
via a two-pronged approach by performing full numerical MFT calculations and by applying the DA beyond MFT.
The corresponding comparison of theoretical and experimental results revealed on one hand that the critical Casimir 
effect is very sensitive to the details of the imprinted chemical substrate 
structures and on the other hand that it can resolve them \cite{troendle:2009,troendle:2011}.

Motivated by these studies, here we investigate the critical Casimir interaction between a \emph{cylindrical}
colloidal particle and a chemically striped substrate.
Highly elliptical and (sphero-) cylindrical colloids are widely used in present research 
and may have applications in new materials \cite{Viry2010,Jiang2011}.
A diverse set of elongated particles such as cylindrical micelles \cite{Gilroy2010}, 
block copolymers \cite{Walther2009}, the mosaic tobacco virus \cite{Namba1986}, 
and carbon nanotubes \cite{Zhao2006, Perez-Juste2005} is experimentally available. 
The (self-) alignment of the latter type of particles can also be used in biological cell 
analyses \cite{Kim2010}, and may lead to improvements in different technologies such as 
solar cells \cite{Yang2011}, energy storage \cite{Reddy2012}, or liquid crystal displays \cite{Russell2006}.
In the context of critical Casimir forces such elongated particles have so far only been investigated in two theoretical studies. In the presence
of a homogeneous substrate MFT predicts the occurrence of a critical
Casimir torque within the plane normal to the substrate surface \cite{kondrat:2009}. 
In Ref.~\cite{Vasilyev2013}, for needle-shaped particles in a two dimensional Ising model such torques have been found 
by conformal field theory and MC simulations.

Besides having an anisotropic shape, particles may also exhibit inhomogeneous surface properties, e.g., Janus particles with two different sides or patchy particles with several surface patches. The experimental fabrication of such particles is of research interest in itself \cite{Prasad2009,Walther2009,Yi:2013}, as is the theoretical understanding of the interactions between spherical \cite{Sciortino2009, Hong2006} or between non-spherical Janus particles \cite{Li2011, Liu2011}, because they are considered to be promising building-blocks for self-assembling materials.

In Ref.~\cite{troendle:2010} an \emph{infinitely} extended cylinder, opposite to and with its horizontal axis aligned parallel 
to the direction of chemical stripes on a surface, has been studied within MFT and within DA.
Here, we extend this previous investigation to cylindrical colloids of \emph{finite} length, which, in addition,
may be arbitrarily \emph{rotated} with respect to the direction of the substrate pattern [\fref{fig:cyl_substrate}].
Thus, the cylindrical colloids are exposed to a critical Casimir torque within the plane parallel
to the substrate surface which tends to align the colloids in the case of a striped substrate.
Adding this orientational degree of freedom enriches the phenomena considerably.
In order to keep complexity in check, we thus focus on the case that the symmetry axis of the cylindrical colloid
remains parallel to the substrate surface.
The critical Casimir torque leading to a tilt out of this plane has been studied for a homogeneous substrate in Ref.~\cite{kondrat:2009}.

As a starting point, in Sec.~\ref{sec:homog} we study a cylindrical colloid of radius $R$ aligned parallel to a chemically 
homogeneous substrate at surface-to-surface distance $D$.
This goes beyond previous analyses in that this colloid is considered to have a \emph{finite} length $L$.
We investigate various combinations $(a,b)$ for the BCs $(a)$ at the substrate and $(b)$ at 
the colloidal surface, in particular for the experimentally relevant  $\{(a),(b)\}\in(\pm)$ BCs for 
binary liquid mixtures.
Whereas $(-,-)$ BCs correspond to the same (strong) adsorption preference [\fref{fig:cyl_substrate}(a)], $(+,-)$ BCs correspond
to opposite adsorption preferences [\fref{fig:cyl_substrate}(b)]. 
Since we are exclusively dealing with a vanishing bulk field, i.e., the concentration of the fluid is kept 
fixed at its critical value, the cases $(+,+)$ and $(-,-)$ as well as the cases $(+,-)$ and $(-,+)$, respectively, are equivalent. 

In Sec.~\ref{sec:step} we consider a cylinder exhibiting a $(-)$ BC close to a chemical step at 
which the adsorption preference 
of the substrate changes discontinuously along the lateral direction $x$ from $(+)$ BC to $(-)$ BC at $x=0$ [\fref{fig:cyl_substrate}(c)]. 
(Experimentally, chemical steps are broadened; we consider the realistic case that the transition region of the chemical surface
composition is narrow on the scale of the diverging bulk correlation length \cite{troendle:2011}.)
The cylinder axis is rotated in the plane parallel to the substrate by an angle $\alpha$ with respect to the chemical step.
We calculate the critical Casimir force, potential, and torque acting on the colloid with its center located at
lateral position $X$.
Based on these results, in Sec.~\ref{sec:compass} we study a Janus cylinder which imposes an inhomogeneous BC 
to the fluid close to a substrate endowed with a chemical pattern of parallel stripes with laterally alternating
BCs [\fref{fig:cyl_substrate}(d)].

In this study we employ a multi-pronged approach. For the homogeneous geometry studied in Sec.~\ref{sec:homog}
we perform full, numerical MFT calculations of the critical Casimir force for a wide range of distance-to-radius
ratios $\Delta\equiv D/R$ and compare them with the corresponding results obtained by using the DA  which is formally 
valid only in the limit $\Delta\to0$. 
However, our analysis reveals that the use of the DA also provides a quantitatively reliable approximation
for experimentally relevant ranges of nonzero values of $\Delta$, so that we can base
our analyses in Secs.~\ref{sec:step} and \ref{sec:compass} on the DA.
Finally, in Sec.~\ref{sec:summary}, we summarize our main findings. 

First, however, in Sec.~\ref{sec:theory} we briefly recall the necessary theoretical background which 
encompasses the appropriate finite size scaling of
critical phenomena and the corresponding field-theoretic approach as well as details of the DA.
\section{Theoretical background \label{sec:theory}}
%
\subsection{Finite-size scaling}
%
In the following we focus on 
binary liquid mixtures upon approaching their critical demixing point by varying the temperature
at constant pressure and with their composition fixed at the critical value \cite{soyka:2008,troendle:2011}.
According to finite-size scaling theory, the critical Casimir force exhibits a universal
behavior described in terms of scaling functions.
These scaling functions depend only on (i) the bulk universality class, which is of Ising type, 
(ii) the surface universality class, which here is of the so-called normal kind and corresponds to either
$(+)$ or $(-)$ BC captured by strong, symmetry-breaking surface fields, (iii) the spatial dimension $d$, which here will be  $d=3$ or $d=4$, and (iv)
the geometry under consideration \cite{krech:book,brankov:book}.
Critical phenomena are to large extent independent of the molecular character of the system due to the divergence of the bulk 
correlation length $\xi_\pm(t\to0^\pm)=\xi_0^\pm|t|^{-\nu}$, where $\nu\simeq0.63$  in $d=3$ and $\nu=1/2$ 
in $d=4$ \cite{pelissetto:2002}, and $t=\pm(T-T_c)/T_c$ is the reduced temperature.
The sign of $t$ is chosen such that $t>0$ corresponds to the homogeneous, mixed phase of the 
fluid, whereas $t<0$ corresponds to the inhomogeneous ordered phase, where phase separation occurs.
For an upper critical point the homogeneous phase is found at high temperatures, and one has $t=(T-T_c)/T_c$.
However, many experimentally relevant binary liquid mixtures, such as the one used in the aforementioned
experiments \cite{hertlein:2008,soyka:2008,gambassi:2009,troendle:2011}, exhibit a lower critical point, 
so that in this case $t=-(T-T_c)/T_c$.
\par
For the \emph{film} geometry, which corresponds to a fluid confined between two parallel, 
infinitely extended walls at distance $l$
the critical Casimir force $f_{(a,b)}$
per area acting on the walls is given by \cite{krech:9192all}
\small
\begin{equation} 
  \label{eq:planar-force}
  f_{(a,b)}(l,T)=k_BT \frac{1}{l^d}k_{(a,b)}( \sgn(t)\, l/\xi_\pm).
\end{equation} 
\normalsize
In \eref{eq:planar-force} the subscript $(a,b)$ denotes the pair of BCs $(a)$ and $(b)$. 
The scaling function $k_{(a,b)}$ depends only on a single scaling variable, which is determined by
the sign of the reduced temperature $t$ and the film thickness $l$ in units of the bulk correlation length $\xi_\pm$
($\pm$ for $t\gtrless0$).
We note that, in general, $(a)$ and $(b)$ can also represent the various symmetry preserving fixed-point BCs 
(the so-called ordinary, special, periodic, or anti-periodic BCs \cite{krech:book,brankov:book}) in addition to 
the symmetry breaking $(\pm)$ BCs relevant to classical one-component fluids or binary liquid mixtures, which we are mainly interested in here.
Right at the bulk critical point the scaling function $\kab$ attains a universal 
constant value $\kab(l/\xi_\pm=0)=\Dab$, known as the critical Casimir amplitude, and the 
critical Casimir force decays algebraically $\sim l^{-d}$ as a function of the film thickness  
\cite{krech:book,brankov:book}.
In contrast, away from criticality, the critical Casimir force decays exponentially as 
a function of $l/\xi_\pm$. 
For $(-,-)$ or $(+,-)$ BCs, which are the experimentally relevant cases \cite{hertlein:2008,soyka:2008,gambassi:2009,troendle:2011},
\emph{and} for the homogeneous phase at $t>0$, we
expect for ${l/\xi_+\gg1}$ (see Refs.~\cite{troendle:2009,troendle:2010})
\small
\begin{equation} 
  \label{eq:exponential-decay}
  k_{(+,\pm)}(l/\xi_+\gg1)=\mathcal{A}_\pm \left(\frac{l}{\xi_+}\right)^d \exp(-l/\xi_+),
\end{equation}
\normalsize
where $\mathcal{A}_\pm$ are universal constants \cite{gambassi:2009}.
\vspace*{-1.2em}
\subsection{Mean-field theory \label{sec:MFT}}
%
Within the field-theoretical approach to critical phenomena, the standard Landau-Ginzburg-Wilson 
bulk Hamiltonian is given by \cite{binder:1983,diehl:1986}
\small
\begin{equation} 
  \label{eq:hamiltonian}
   \mathcal{H}[\phi]=\int_V\,\upd^d\vec{r}\,\left\{
        \frac{1}{2}(\nabla\phi)^2
       +\frac{\tau}{2}\phi^2
       +\frac{u}{4!}\phi^4
			 \right\},
\end{equation} 
\normalsize
where $\phi(\vec{r},t)$ is the spatially varying and temperature dependent
order parameter describing the binary
fluid, i.e., the difference between the local concentration of one of the two species and its critical
value.
The fluid completely fills the volume $V$ in $d$-dimensional space.
The parameter $\tau$ in \eref{eq:hamiltonian} is proportional to the reduced temperature $t$, 
and the coupling constant $u>0$ provides stability of the Hamiltonian for $t<0$.
The mean-field order parameter profile $m(\vec{r},\tau)\ddef u^{1/2}\langle\phi(\vec{r},t)\rangle$ is
defined as the one which minimizes the Hamiltonian, i.e., 
$\updelta \mathcal{H}[\phi]/\updelta\phi|_{\phi=\langle\phi\rangle}=0$ for given BCs.
Within renormalization group theory, MFT is the lowest order contribution within an expansion in terms of 
$\varepsilon=4-d$, and it allows one to infer the universal scaling 
functions of the critical Casimir force at the upper critical dimension $d=4$ (up to an overall 
prefactor $\propto u^{-1}$ and up to logarithmic corrections). 
Thus, in the following, we denote MFT results by $d=4$, in contrast to results obtained from
MC simulation data in physical dimensions $d=3$ via the DA (see below).
The \emph{bulk} mean-field order parameter is given by $\langle\phi\rangle=\pm a|t|^\beta$ for $t<0$, where
$\beta$ is a standard critical exponent, and it vanishes for $t>0$. 
The non-universal amplitudes $\xi_0^+$ and $a$ are the only two independent non-universal amplitudes
appearing in the description of bulk critical phenomena (two-scale universality) \cite{binder:1983,diehl:1986}, 
so that, within MFT ($d=4$) the relationships $\tau=t (\xi_0^+)^{-2}$, $u=6(a\xi_0^+)^{-2}$, $R_\xi\equiv\xi_0^+/\xi_0^-=\sqrt{2}$,
and $m_{\textrm{bulk}}=\sqrt{6}(\xi_0^+)^{-1}|t|^{1/2}$
hold \cite{tarko:all}.
In a \emph{confined} system, $\mathcal{H}[\phi]$ is supplemented by appropriate surface contributions 
\cite{binder:1983,diehl:1986}, which for the normal universality class \cite{burkhardt:1994,diehl:1993}
generate the fixed-point $(\pm)$ BCs for the order parameter such that $\phi\big|_{\text{surface}}=\pm\infty$.
\par
For the \emph{film} geometry the order parameter profiles $m(\vec{r},\tau)$ which minimize $\mathcal{H}[\phi]$ 
and the resulting critical Casimir forces for the various $(a,b)$ BCs have been determined analytically \cite{krech:1997}.
For the case of the symmetry breaking BCs, the critical Casimir amplitudes are given by
$\Dpp=\Dmm=-\Dpm/4=48[K(1/\sqrt{2})]^4 u^{-1}$
where $K$ is the complete 
elliptic integral of the first kind \cite{krech:1997}.
\par
 
Here, we study a cylindrical colloid near a planar substrate, as shown in \fref{fig:cyl_substrate}.
In order to obtain the MFT order parameter profile $m(\vec{r},\tau)$ for such geometries, we have 
minimized \eref{eq:hamiltonian} numerically for various BCs using a finite element method \cite{kondrat:f3dm}.
The fixed point $(\pm)$ BCs have been implemented numerically by means of a short distance expansion 
at the planar and curved surfaces \cite{hanke:1999a,kondrat:2007}.
Subsequently, we have calculated the critical Casimir force acting on the cylindrical colloid from these 
profiles using the stress tensor \cite{krech:1997,kondrat:2009,bitbol:2011}.
This renders the universal scaling function for the critical Casimir force in ${d=4}$, in which the
surface of the cylinder is still bent only along one direction as in $d=3$, but all physical properties 
are  invariant along the additional fourth dimension. 
Accordingly, the MFT results are those per length along this fourth direction.
We estimate the relative numerical error of the MFT scaling functions obtained this way to be less than 1\%
for the case $L\to\infty$ presented in Sec.~\ref{sec:infinite} below, 
and less than 3\% for finite values of $L$ presented in Sec.~\ref{sec:finite}.
\vspace*{-1.2em}
\subsection{Derjaguin approximation \label{sec:da}}
%
Since soft matter naturally involves curved objects, the well-known DA is often used
in order to deal with spherical surfaces \cite{derjaguin:1934}.
Accordingly, the critical Casimir force acting on a spherical colloid in the proximity of a planar substrate
can be calculated via subdividing the spherical surface into infinitely small rings parallel to the substrate 
surface.
Using the known expression for the force in the plane-plane geometry and assuming additivity, the overall force
acting on the colloid can be obtained by simply summing up the individual contributions from these rings 
\cite{hanke:1998}. 
The DA has been used successfully in order to predict the
experimentally measured critical Casimir force acting on spherical colloids \cite{hertlein:2008,gambassi:2009,
troendle:2009,troendle:2011}.

In this study, we apply the DA to cylindrical particles \cite{troendle:2010}, such that
the surface of the cylindrical colloid is decomposed into pairs of infinitely narrow stripes of length $L$
positioned parallely to the cylinder axis and the substrate.
The distance of each stripe from the substrate surface is given by $D(\rho)=D+R(1-\sqrt{1-\rho^2/R^2})$,
where
$\rho$ is the absolute value of the position of each stripe along the lateral 
$x$ direction.
The DA holds in the limit $\Delta=D/R\to0$ so that here we use the ``parabolic distance approximation''
 \cite{hanke:1998,troendle:2010}
\small
\begin{equation}
  \label{eq:parabolic}
  D(\rho)\simeq D\left(1+\rho^2/(2RD)\right).
\end{equation}
\normalsize

Assuming additivity of the forces and neglecting edge effects, we subsequently integrate contributions 
from the single stripes forming the cylinder surface by using the expression for the force in the film geometry, as 
given by \eref{eq:planar-force}, for the appropriate distance $l=D(\rho)$.
For the cases shown in Fig.~\ref{fig:cyl_substrate}(a) and (b) the BCs of the 
substrate and the cylinder are spatially homogeneous, so that the corresponding universal scaling function 
for the film geometry is either $\kmm$ or $\kpm$.
For the more complex geometries shown in Fig.~\ref{fig:cyl_substrate}(c) and (d) the cylinder faces
a spatially varying adsorption preference at the substrate. 
Thus, in this case the summation of forces demands a further subdivision of the cylinder surfaces 
depending on the projection of the cylinder onto the substrate surface, 
analogously to the case of a sphere near a chemically patterned substrate \cite{troendle:2010},
and as described in detail in Appendix~\ref{sec:app-da} and in Ref.~\cite{labbe:thesis}.
Accordingly, for the cases shown in Figs.~\ref{fig:cyl_substrate}(c) and (d), the integral over force contributions
depends on both scaling functions $\kmm$ and $\kpm$.

We note that, within the DA, we only take into account forces which are directed normal to the substrate surface,
motivated by the studies of forces between a homogeneous particle and a homogeneous substrate, 
for which it turns out that local stresses acting on the particle surface along lateral 
directions in sum cancel out.
Nonetheless, lateral forces acting on the colloid close to a patterned substrate are also obtained via this approach by 
constructing an effective interaction potential which depends on the lateral coordinate and is 
derived via integrating the normal force.
Other approaches consider force contributions acting locally normal to the surface of the colloidal 
particle from the very beginning \cite{dantchev:2012}. 
However, in the case of a sphere close to a homogeneous substrate the different approaches 
tend to the same expression, taking into account the different nature of body forces, as considered in Ref.~\cite{dantchev:2012}, and critical Casimir forces, which only act on surfaces.

The DA is based on the universal scaling functions for the film geometry, which
have been obtained from MC simulations in $d=3$ \cite{vasilyev:2007,vasilyev:2009,
dantchev:2004,hasenbusch:2010a,hasenbusch:2012}.
In the remainder of our study, concerning the application of the DA in $d=3$ we use for the scaling functions $k_{(\pm,-)}$ 
the numerical estimate referred to as ``approximation (\textit{i})'' in Figs.~9 and 10 of Ref.~\cite{vasilyev:2009}. 
The systematic uncertainty of the overall amplitude of these scaling functions can, in the worst case, reach up to 10\%--20\% 
\cite{vasilyev:2009}, which also affects our predictions.
However, we have checked that the corresponding impact on the \emph{normalized} scaling functions 
presented below is much smaller and only on the relative level of at most $5\%$.
The same estimate also applies to the use of more recent MC data \cite{hasenbusch:2010a,hasenbusch:2012}.

\vspace*{-1.5em}
\section{Cylinder close to a homogeneous substrate\label{sec:homog}}
%
In order to predict the universal scaling function for the critical Casimir forces acting
on a cylindrical colloid in $d=3$ we employ the DA discussed
in the preceding section.
Recently, the DA has been compared with data obtained by MC simulations in $d=3$ for 
the sphere-wall geometry \cite{hasenbusch:2013}, showing that the accuracy of the DA depends on 
the type of BCs.
Whereas for $(+,+)$ BCs (equivalent to $(-,-)$ BCs), the DA agrees rather well with the MC 
data for colloid-substrate distances of up to roughly one particle radius, for $(+,-)$ BCs a notable 
disagreement has been found for these geometries and for $t<0$ \cite{hasenbusch:2013}.
The aim of this section is to elucidate this issue in more detail by inspecting the range of validity
of the DA for the critical Casimir force acting on a cylindrical particle.

To this end, we consider a cylindrical colloid of radius $R$, length $L$, and with $(b)$ BC 
close and parallel to a chemically homogeneous, flat wall with $(a)$ BC and surface-to-surface 
distance $D$.
In particular, we are interested in $(a=\pm,b=-)$ BCs, as shown in
Figs.~\ref{fig:cyl_substrate}(a) and (b).

According to finite size scaling, the normal critical Casimir force $\Fab(D,R,L,T)$ and the critical Casimir 
potential $\Phiab(D,R,L,T)\equiv\int_D^\infty \upd z\ \Fab(z,R,L,T)$ acting on the colloid
can be written as \cite{troendle:2010}
\small
\begin{equation} 
  \label{eq:cyl-force-homog}
  \Fab(D,R,L,T)=k_BT\frac{LR^{1/2}}{D^{d-1/2}}\Kab(\Theta,\Delta,\Len),
\end{equation} 
\normalsize
and
\small
\begin{equation} 
  \label{eq:cyl-pot-homog}
  \Phiab(D,R,L,T)=k_BT\frac{LR^{1/2}}{D^{d-3/2}}\varthetaab(\Theta,\Delta,\Len),
\end{equation} 
\normalsize
respectively.
In Eqs.~\eqref{eq:cyl-force-homog} and \eqref{eq:cyl-pot-homog} $\Kab$ and $\varthetaab$ are universal scaling functions
which only depend on the dimensionless scaling variables 
\small
\begin{equation}
  \label{eq:scalingvariables}
  \Theta\equiv\sgn(t)\frac{D}{\xi_\pm},
  \qquad\Delta\equiv \frac{D}{R},
  \qquad\Len\equiv \frac{L}{\sqrt{RD}},
\end{equation}
\normalsize
i.e., $\Theta=D/\xi_+$ for $t>0$ and $\Theta=-D/\xi_-$ for $t<0$.
Note that using $\Theta$ is equivalent to using the scaling variable
\small
\begin{equation} 
  \label{eq:y}
  \hat{y}\equiv t \left(\frac{D}{\xi_0^+}\right)^{\frac{1}{\nu}}=\left\{
  \begin{split}
    \Theta^{1/\nu}, \qquad t>0,\\
    -(|\Theta|/R_\xi)^{1/\nu}, \;\;t<0,\\
  \end{split}
  \right.
\end{equation}
\normalsize
which is widely used for the film geometry.
Eq.~\eqref{eq:cyl-force-homog} describes the force acting on the cylinder per [length]$^{d-3}$ which for $d=3$ corresponds to the force and for
$d=4$ corresponds to $\Fab$ per length of the extra translationally invariant direction.
According to $\Phiab(D,R,L,T)\equiv\int_D^\infty \upd z\ \Fab(z,R,L,T)$, the scaling 
function $\varthetaab(\Theta,\Delta,\Len)$ in Eq.~\eqref{eq:cyl-pot-homog} can be expressed in terms of the scaling function 
$\Kab(\Theta,\Delta,\Len)$ of the force [\eref{eq:cyl-force-homog}]:
\small
\begin{equation}
\varthetaab(\Theta,\Delta,\Len)= \int\limits_1^\infty \upd z\ \frac{\Kab(z\, \Theta,  z\, \Delta,\Len)}{{z}^{d-1/2}}.
\label{eq:scalefunc_cyl_pot_relation}
\end{equation}
\normalsize
Conversely, using $\Fab(D,R,L,T)=-\frac{\upd}{\upd z}\Phiab(z,R,L,T)\rvert_{z=D}$, the scaling function of the critical Casimir force $\Kab(\Theta,\Delta,\Len)$ is given in terms of $\varthetaab(\Theta,\Delta,\Len)$ by the total derivative
\small
\begin{align}
\Kab(\Theta, \Delta,\Len) = &
(d-3/2) \varthetaab(\Theta,\Delta,\Len)\nonumber\\
  & - \Theta\,\frac{\partial}{\partial\Theta} \varthetaab(\Theta,\Delta,\Len) \nonumber \\
  & - \Delta\,\frac{\partial}{\partial \Delta}\varthetaab(\Theta,\Delta,\Len).
\label{eq:scalefunc_cyl_force_relation}
\end{align}
\normalsize

\vspace*{-1.2em}
\subsection{Derjaguin Approximation\label{sec:homog-da}}
%
%
First, we consider the scaling functions for the critical Casimir force and the corresponding potential
within the DA (see Sec.~\ref{sec:da}), which corresponds to the limit $\Delta\to0$.
We obtain for the scaling function of the critical 
Casimir force (see also Appendix D of Ref.~\cite{troendle:2010})
\small
\begin{equation} 
  \label{eq:cyl-force-da}
  \Kab(\Theta,\Delta\to0,\Len)=\sqrt{2}\int\limits_1^\infty\upd\beta\,(\beta-1)^{-\frac{1}{2}}\,\beta^{-d}\,\kab(\Theta\beta).
\end{equation} 
\normalsize
Right at the bulk critical point $\Theta=0$ we find $\Kab(0,0,\Len)=\sqrt{2\pi}[\Gamma(d-\frac{1}{2})/\Gamma(d)]\Dab$ 
so that $\Kab(0,0,\Len)=[3\pi/(4\sqrt{2})]\Dab $ for $d=3$ and 
$\Kab(0,0,\Len)=[5\pi/(8\sqrt{2})]\Dab $ for $d=4$ \cite{troendle:2010}.

Accordingly, the scaling function of the critical Casimir potential can be found by integrating over the critical Casimir 
force using Eqs.~\eqref{eq:scalefunc_cyl_pot_relation} and \eqref{eq:cyl-force-da}, so that \cite{labbe:thesis}
\ifTwocolumn
\small
\begin{align} 
  \varthetaab&(\Theta,\Delta\to0,\Len)= \nonumber\\
  &\sqrt{2}\int\limits_1^\infty \upd z\ 
\int\limits_1^\infty\upd\beta\,(\beta-1)^{-\frac{1}{2}}\,\beta^{-d}\,{z}^{-d+1/2}\,\kab(\Theta\beta z)\nonumber\\
=&2\sqrt{2}\int\limits_1^\infty\upd\nu\,\sqrt{\nu-1}\,\nu^{-d}\,\kab(\Theta\nu),
\label{eq:cyl-pot-da}
\end{align}
\normalsize
\else
\begin{eqnarray} 
  \varthetaab(\Theta,\Delta\to0,\Len)&=&
\sqrt{2}\int\limits_1^\infty \upd z\ 
\int\limits_1^\infty\upd\beta\,(\beta-1)^{-\frac{1}{2}}\,\beta^{-d}\,{z}^{-d+1/2}\,\kab(\Theta\beta z)\nonumber
  \\&=&2\sqrt{2}\int\limits_1^\infty\upd\nu\,\sqrt{\nu-1}\,
  \nu^{-d}\,\kab(\Theta\nu),
  \label{eq:cyl-pot-da}
\end{eqnarray} 
\fi
where we have changed the variable $\beta \text{ into } \nu \equiv \beta z$, exchanged the order of the remaining integrals 
$\int_1^\infty \upd z \int_z^\infty\upd\nu = \int_1^\infty\upd\nu \int_1^\nu \upd z$, and used 
$\int_1^\nu \upd z (\nu-z)^{-1/2} = 2 \sqrt{\nu-1}$.
For $\Theta=0$, $\varthetaab(0,0,\Len)=\sqrt{2\pi}[\Gamma(d-\frac{3}{2})/\Gamma(d)]\Dab$ 
so that $\varthetaab(0,0,\Len)=[\pi/(2\sqrt{2})]\Dab  $ for $d=3$ and 
$\varthetaab(0,0,\Len)=[\pi/(4\sqrt{2})]\Dab $ for $d=4$.
\par
Note that the expressions for both $\Kab$ and $\varthetaab$ given in Eqs.~\eqref{eq:cyl-force-da} 
and \eqref{eq:cyl-pot-da} are independent of the value of $\Len$ because the original DA neglects edge effects.
Thus, within the DA, the dependence on the length of the cylinder reduces simply to an overall prefactor $L$
in Eqs.~\eqref{eq:cyl-force-homog} and \eqref{eq:cyl-pot-homog}.
\vspace*{-1.2em}
\subsection{MFT scaling functions for an infinitely extended cylinder\label{sec:infinite}}
%
\begin{figure}[b!]
  \begin{center}
   \ifTwocolumn
  \includegraphics[width=7.5cm]{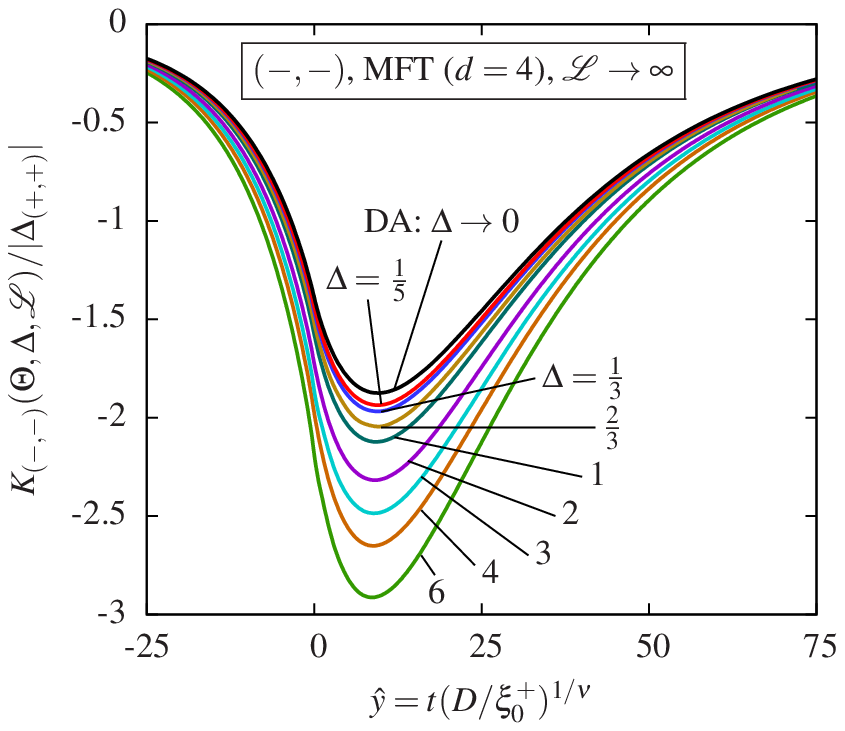}
  \else
  \includegraphics[width=11cm]{fig_02}
  \fi
\end{center}
  \clevercaption{%
  Reduced MFT scaling function $K_{(-,-)}(\Theta,\Delta,\Len)/|\Dpp|$ [\eref{eq:cyl-force-homog}] for 
  the attractive critical Casimir force acting on a cylindrical particle near a planar substrate, both sharing
  the same $(-)$ BC at their surfaces (see \fref{fig:cyl_substrate}(a)).
  The numerically obtained universal scaling function $K_{(-,-)}$ is shown as a function
  of $\hat{y}=t(D/\xi_0^+)^{1/\nu}$ for an infinitely extended cylinder 
  with $\Len=L/\sqrt{RD}\to\infty$ and various values of $\Delta=D/R$ ranging from $1/5$ to $6$ for temperatures above and below $T_c$.
  With decreasing values of $\Delta$, the scaling function uniformly approaches the corresponding
  one obtained within the DA [\eref{eq:cyl-force-da}], which is valid in the limit $\Delta\to0$.
    }   
  \label{fig:homog_pp}
\end{figure}

In order to test the range of validity of the DA, which is determined by the geometrical parameters of the 
system under consideration, we have numerically calculated the full MFT scaling functions for a wide
range of values of $\Delta=D/R$ and for $\Len\to\infty$.

First, we focus on the attractive interaction case of $(-,-)$ BCs as shown in \fref{fig:cyl_substrate}(a).
Figure~\ref{fig:homog_pp} shows the reduced MFT scaling functions 
$\Kmm(\Theta,\Delta,\Len\to\infty)/|\Dpp|$ in terms of the absolute value of the Casimir amplitude 
$\Dpp$ for the film geometry as a function of $\hat{y}$ (see \eref{eq:y}).
Moreover, we compare these results with the DA obtained for $\Delta\to0$.
From \fref{fig:homog_pp} we can infer that the DA limit is approached uniformly upon decreasing
the distance-to-radius ratio $\Delta$.
In fact, the DA results agree quantitatively rather well with the full,
numerical scaling functions for $\Delta\lesssim 1/3$.
We note that the DA underestimates the strength of the force as compared with the 
actual data for all values of $\hat{y}$ and $\Delta$.
This is similar to the case of a sphere near a planar substrate and $(+,+)$ BCs as
has been discussed for MFT in Ref.~\cite{troendle:2010} and for $d=3$ by using
MC simulation data in Ref.~\cite{hasenbusch:2013}.
Thus, in the case of a cylinder
near a planar substrate we expect this behavior to carry over accordingly to $d=3$.

\begin{figure}[b!]
  \begin{center}
   \ifTwocolumn
  \includegraphics[width=7.5cm]{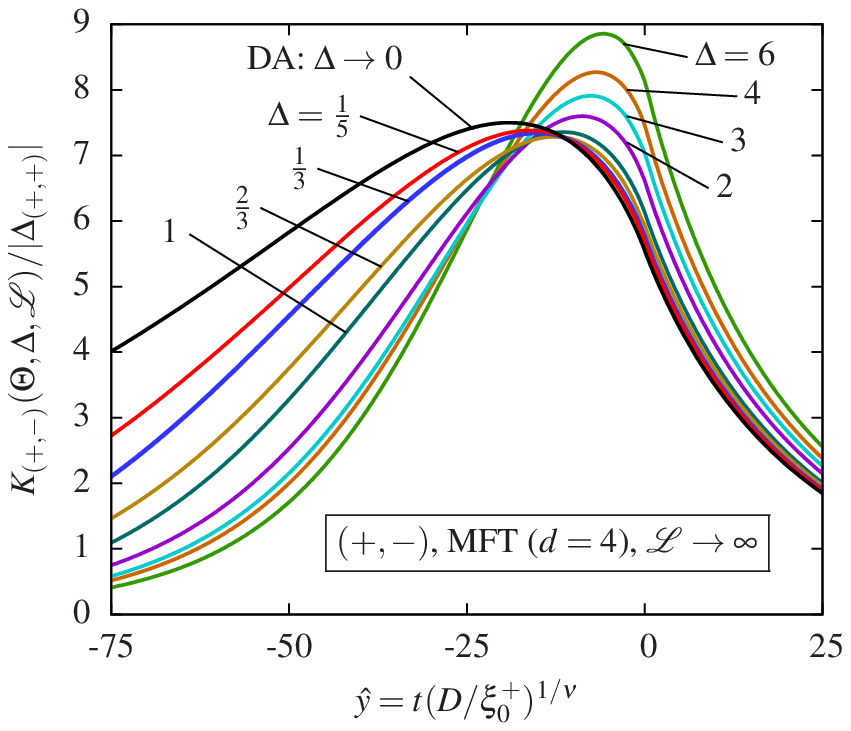}
  \else
  \includegraphics[width=11cm]{fig_03}
  \fi
\end{center}
  \clevercaption{%
  Same as \fref{fig:homog_pp}, but for opposite $(+,-)$ BCs at the surfaces of the colloid
  and the substrate (see \fref{fig:cyl_substrate}(b)), resulting
  in a repulsive critical Casimir force described by the  reduced scaling function 
  $\Kpm(\Theta,\Delta,\Len\to\infty)/|\Dpp|$. 
  For $\hat{y}\ge0$, as for $(-,-)$ BCs shown in \fref{fig:homog_pp}, the scaling
  function uniformly approaches the DA limit $\Delta\to0$, and is
  in good agreement with it for $\Delta\lesssim1/3$.
  However, for $\hat{y}<0$ the discrepancies between the scaling function as obtained
  within the DA and the full, numerically obtained one are much larger.
  This rich behavior of the scaling function as function of $\Delta$ resembles the one
  obtained from MC simulations for a spherical particle close to a planar substrate \cite{hasenbusch:2013}.
    }   
  \label{fig:homog_pm}
  \ifTwocolumn\vspace{-1em}\fi
\end{figure}

For opposite $(+,-)$ BCs at the colloid and the substrate surfaces (\fref{fig:cyl_substrate}(b)) 
the corresponding behavior of the scaling functions for the critical Casimir force, which in this case is repulsive,
is more involved (see \fref{fig:homog_pm}).
Whereas for $t\ge0$ (i.e., $\hat{y}\ge0$) the reduced scaling functions $\Kpm(\Theta,\Delta,\Len\to\infty)/|\Dpp|$
uniformly approach the DA limit for $\Delta\to0$ from above, this behavior drastically changes for negative values
of $\hat{y}$.
As can be inferred from \fref{fig:homog_pm}, for $\hat{y}\ll-1$, depending on $\Delta$, the DA  
overestimates the actual force. 
Moreover, the DA limit is reached only for much smaller values of $\Delta$ as compared with the case $\hat{y}\ge0$.
Indeed, even for $\Delta=1/5$ the full MFT scaling function differs significantly from the DA
for $\hat{y}\ll-1$.
This pronounced dependence on the value of $\Delta$ has also been observed in $d=3$ for the sphere-wall geometry
and $(+,-)$ BCs, for which the corresponding scaling functions obtained from MC simulations show a
qualitatively similar behavior \cite{hasenbusch:2013}.

\begin{figure} 
  \begin{center}
   \ifTwocolumn
  \includegraphics[width=7.5cm]{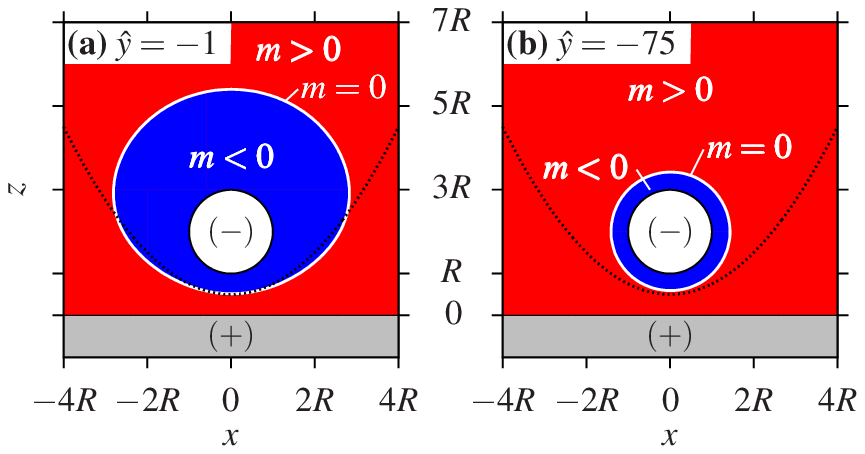}\\
  \includegraphics[width=7.5cm]{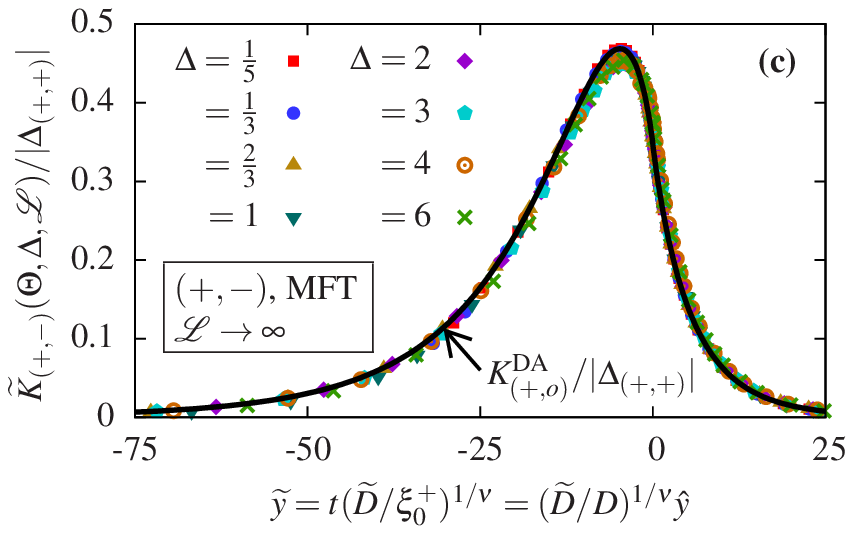}
  \else
  \includegraphics[width=10cm]{fig_04ab}\\
  \includegraphics[width=10cm]{fig_04c}
  \fi
\end{center}
  \clevercaption{%
   \ifTwocolumn
   \else
\setstretch{1.2}
\fi
  {(a,b):} Location of the interface manifold $m=0$ in a binary liquid, forming for $t<0$ between the two
  phases rich in either one of the constituents. The phases $m<0$ (blue) and $m>0$ (red) are preferred by the colloid (white) with
  $(-)$ BC and the substrate (grey) exhibiting $(+)$ BC, respectively (see \fref{fig:cyl_substrate}(b)).
  The axis of the cylindrical colloid is parallel to the substrate and located at $(x=0,z=2R)$ so that the system is translationally invariant in the direction
  normal to the $xz$ plane. Accordingly, the vertical projection $z_0(x)$ of the interface manifold onto the $xz$ plane is shown as solid white line for
  the specific example $\Delta=1$ (i.e., $D=R$) and $\hat{y}=-1$ in (a) and $\hat{y}=-75$ in (b).
  Whereas for $\hat{y}=-1$ in (a) the interface is, over a broad range of values of $x$, located at $z_0(x=0)=D/2$ in the region of closest approach
  between the colloid and the substrate, for $\hat{y}=-75$ in (b) the interface is closely wrapped around
  the colloid.
  Thus, for $\hat{y}=-1$ in (a) for a wide range of values of $x$ the actual interface location around $x=0$ is in accordance with the 
  DA assumption   $z_0^{\textrm{DA}}(x)$ (see the main text) shown as a dotted black line, but they disagree
  for $\hat{y}=-75$ in (b), i.e., the dotted black line does not fall on top of the solid white line even for $x=0$.
  {(c):} Reduced effective scaling functions $\widetilde{K}_{(+,-)}(\Theta,\Delta,\Len)/|\Dpp|$
  (see \eref{eq:Ktilde}) as a function of $\widetilde{y}=t(\widetilde{D}/\xi_0^+)^{1/\nu}=(\widetilde{D}/D)^{1/\nu}\hat{y}$, where
  $\widetilde{D}$ is the distance $z_0(x=0)$ of the interface $m=0$ from the substrate.
  The symbols correspond to the full MFT data for $(+,-)$ BCs shown in \fref{fig:homog_pm}
  but rescaled, accounting for the interface shape sketched in (a) and (b) (see the
  main text).
  Accordingly, for various values of $\Delta$ within the range from $1/5$ to $6$ the scaling function for the critical
  Casimir force basically collapses onto a single master curve. In addition, the latter is in good agreement
  well with the corresponding scaling function $K_{(+,o)}^{\textrm{DA}}$ which has 
  been obtained within a modified DA on the basis of the interface shape obtained numerically from the full MFT
  for each configuration.
    }   
  \label{fig:homog_mod}
\end{figure}

Based on the order parameter profiles, which we have obtained numerically within MFT, we attribute the 
poorer performance of the DA for $(+,-)$ BCs and $t<0$ to the formation of an interface in this demixed state.
Indeed, in general the shape and the location of the interface formed around the colloidal particle differ significantly
from what is assumed within the DA.
This is shown in Figs.~\ref{fig:homog_mod}(a) and (b), where we illustrate for two examples the location of the interface at which
the MFT order parameter $m(\vec{r},\tau)$ vanishes for the specifically chosen values $\Delta=1$ and 
$\hat{y}=-1$ [\fref{fig:homog_mod}(a)] and $\hat{y}=-75$ [\fref{fig:homog_mod}(b)].
In particular, due to the translational invariance in the $y$ direction, the vertical projection $z_0(x)$ of the interface manifold
$m=0$ onto the $xz$ plane is shown as a white line
separating the regions of positive and negative order parameter.
On the other hand, 
due to the intrinsic symmetry of the $(+,-)$ configuration in the film geometry \cite{krech:1997} close to $T_c$ \footnote{%
Far below $T=T_c$ a spontaneous symmetry breaking of the interface occurs which is known as the interface localization-delocalization transition \cite{Parry:1992,Binder:1996}.
Throughout this study we consider $\hat{y}$ to be not too negative and our system to be sufficiently close to criticality, so that the equilibrium configuration for the film geometry with opposing boundary conditions is symmetric with respect to the midplane of the film.
}, the DA assumes the interface manifold $m=0$ to be 
located at $z_0^{\textrm{DA}}(x)=D(x)/2$ with $D(x)$ as the ``parabolic distance approximation'' from \eref{eq:parabolic}.
\nocite{Parry:1992,Binder:1996}
In Figs.~\ref{fig:homog_mod}(a) and \ref{fig:homog_mod}(b) $z_0^{DA}$ is shown as a dotted black line.
Whereas for $\hat{y}=-1$ in \fref{fig:homog_mod}(a) $z_0^{DA}(x)$ agrees well over a wide range with $z_0(x)$, in particular
in the region of closest approach between the colloid and the substrate around $x=0$ 
(with the contributions to the stress tensor from there dominating the resulting critical Casimir force), the two curves deviate much stronger for $\hat{y}=-75$ in \fref{fig:homog_mod}(b).
Whilst the DA always assumes the same position of the interface independent of $t$ (i.e., the dotted lines in Figs.~\ref{fig:homog_mod}(a) and \ref{fig:homog_mod}(b) are identical), 
in fact, below $T_c$ the interface ``snaps'' around the colloidal particle because it tends to minimize its area due to the positive
interfacial tension involved \cite{law:2013}.
This is in accordance with the behavior of the critical Casimir force as shown in \fref{fig:homog_pm},
where the DA strongly deviates from the scaling function, e.g., for $\Delta=1$ for values $\hat{y}\ll-1$ but is
rather close to this scaling function for $\hat{y}\gtrsim-1$.
Although $z_0(x)$ is always closed for $\hat{y}<0$ in contrast to $z_0^{DA}$ (see the solid white lines and the dotted black lines, respectively, in 
  Figs.~\ref{fig:homog_mod}(a) and \ref{fig:homog_mod}(b)), their agreement in the region of closest approach of the colloid and the wall is sufficient for 
a good performance of the DA for $-1 \lesssim \hat{y} <0$.

In order to check whether the discrepancies between the DA and the actual force for $(+,-)$ BCs are 
indeed due to this behavior of the shape of the interface, we introduce a modified version of the 
DA.
From the numerically minimized order parameter profiles we determine for each geometry and temperature
an effective distance $\widetilde{D}\equiv z_0(x=0)$ (note that $z_0(x=0)$ is close to but not identical to $D/2$) 
and an effective radius $\widetilde{R}\equiv(\partial^2_x z_0(x)|_{x=0})^{-1}$ of the interface $m=0$ at $x=0$.
Subsequently, we define an effective scaling function (compare \eref{eq:cyl-force-homog})
\ifTwocolumn
\small
\begin{multline} 
  \label{eq:Ktilde}
  \widetilde{K}_{(+,-)}(\Theta,\Delta,\Len)\equiv
  \frac{\widetilde{D}^{d-1/2}}{L\widetilde{R}^{1/2}}\frac{\Fpm(D,R,L,T)}{k_BT}\\
  =\sqrt{\frac{R}{\widetilde{R}}\left(\frac{\widetilde{D}}{D}\right)^{2d-1}}\;{K}_{(+,-)}(\Theta,\Delta,\Len),
\end{multline} 
\normalsize
\else
\begin{equation}
  \label{eq:Ktilde}
  \widetilde{K}_{(+,-)}(\Theta,\Delta,\Len)\equiv
  \frac{\widetilde{D}^{d-1/2}}{L\widetilde{R}^{1/2}}\frac{\Fpm(D,R,L,T)}{k_BT}\\
  =\sqrt{\frac{R}{\widetilde{R}}\left(\frac{\widetilde{D}}{D}\right)^{2d-1}}\;{K}_{(+,-)}(\Theta,\Delta,\Len),
\end{equation}
\fi
based on $\widetilde{D}$ and $\widetilde{R}$, which themselves depend on $\Theta$, $\Delta$, and $\Len$ via $z_0(x)$.
Thus, $\widetilde{K}_{(+,-)}$ corresponds to the scaling function for the critical
Casimir force, which would be exerted onto a fictitious cylindrical colloid of length $L$ and of radius $\widetilde{R}$
at a surface-to-surface distance $\widetilde{D}$ from the substrate, which exhibits a 
Dirichlet BC $\phi=0$ at its surface.
Although such an $(o)$ BC does generally not occur in binary liquid mixtures, it is the
appropriate fixed point BC in many other cases, so that the corresponding
film scaling function $k_{(+,o)}$ is well known (see Ref.~\cite{krech:1997} for MFT).
Thus, the DA can be analogously performed according to \eref{eq:cyl-force-da}, but based
upon $k_{(+,o)}$. This leads to the effective force scaling function $K_{(+,o)}^{\textrm{DA}}
\equiv  K_{(+,o)}(\pm\widetilde{D}/\xi_\pm,0,\infty)$
corresponding to the fictitious colloid the surface of which is, up to quadratic terms, given by $z_0(x)$ which determines
$\widetilde{D}$ and $\widetilde{R}$.
The scaling function $\widetilde{K}_{(+,-)}$ as obtained from the full, numerical MFT data
is shown with symbols in \fref{fig:homog_mod}(c) as a function of the effective scaling variable
$\widetilde{y}\equiv(\widetilde{D}/\xi_0^+)^{1/\nu}=(\widetilde{D}/{D})^{1/\nu}\hat{y}$.
These data have been obtained for certain values of $\hat{y}$ and for precisely the same values of $\Delta$ as in \fref{fig:homog_pm} before.
Indeed, the data almost perfectly collapse for all values of $\hat{y}$ even for large values of $\Delta$,
and they agree with the DA expression obtained from \eref{eq:cyl-force-da} for $(+,o)$ BCs.
Our comparison, therefore, shows that the aforementioned deficiencies of the DA results for $(+,-)$ BCs and $t<0$
are solely due to the specific shape of the interface around the colloid.
It turns out that within MFT, this issue can be fully resolved in terms of a modified DA for fictitious
$(+,o)$ BCs if the location and the curvature of the actual interface at $x=0$ are known.
Note, however, that this mapping onto a fictitious ersatz colloid with Dirichlet BC is not expected to hold beyond MFT,
because the intrinsic order parameter profile vanishes at $z_0(x)$ linearly $\propto z-z_0(x)$, whereas at a Dirichlet wall
it vanishes $\propto(z-z_0(x))^{(\beta_1-\beta)/\nu}$, where $\beta_1$ is a surface critical exponent 
($\beta_1(d=4)=1$, $\beta_1(d=3)\simeq0.8$ \cite{diehl:1997}).
Thus, within MFT $(\beta_1-\beta)/\nu$ happens to be equal to $1$, too, whereas in $d=3$ $(\beta_1-\beta)/\nu\simeq0.8$.
Despite these quantitative differences, we expect that the analogous poor performance of the DA for $\hat{y}<1$ and $(+,-)$ BCs in $d=3$ \cite{hasenbusch:2013}
can be attributed to the similar, peculiar shape of the interface in this case, too.

\subsection{MFT scaling functions for a cylinder of finite length \label{sec:finite}}

In this subsection we study the dependence of the critical Casimir force on the length $L$
of the cylinder. 
The finite length of the cylinder is accounted for by the scaling variable $\Len=L/\sqrt{RD}$, so 
that the aspect ratio of the cylinder is given by $L/R=\Len\sqrt{\Delta}$.
In the following we shall focus on small ratios $\Delta=D/R$.
We expect the critical Casimir interaction to be strongly affected by 
the influence of the two planar surfaces at the ends of the cylinder, in particular for ``disk-like'' cylinders with $\Len=L/\sqrt{RD}\lesssim 2$.
On the other hand, for $\Len\gg1$ the relative effect induced by the cylinder ends on the
scaling function of the total force acting on the colloid is expected to vanish as 
$\propto \Len^{-1}$ because for $L\gg R,D$ the corresponding force contributions from the ends do
not depend on $L$, whereas the force on a long cylinder is proportional to $L$. 
Since $L$ has been accounted for in \eref{eq:cyl-force-homog}, we consider the expression
\small
\begin{equation} 
  \label{eq:long}
  K_{(\pm,-)}(\Theta,\Delta,\Len\gg1)\simeq K_{(\pm,-)}(\Theta,\Delta,\Len\to\infty) 
  + \frac{\kappa_\pm|\Dpp|}{\Len},
\end{equation} 
\normalsize
where the generalized amplitudes $\kappa_\pm$ depend on $\Theta$ and $\Delta$.

\begin{figure}[t!]
  \begin{center}
   \ifTwocolumn
  \includegraphics[width=7.5cm]{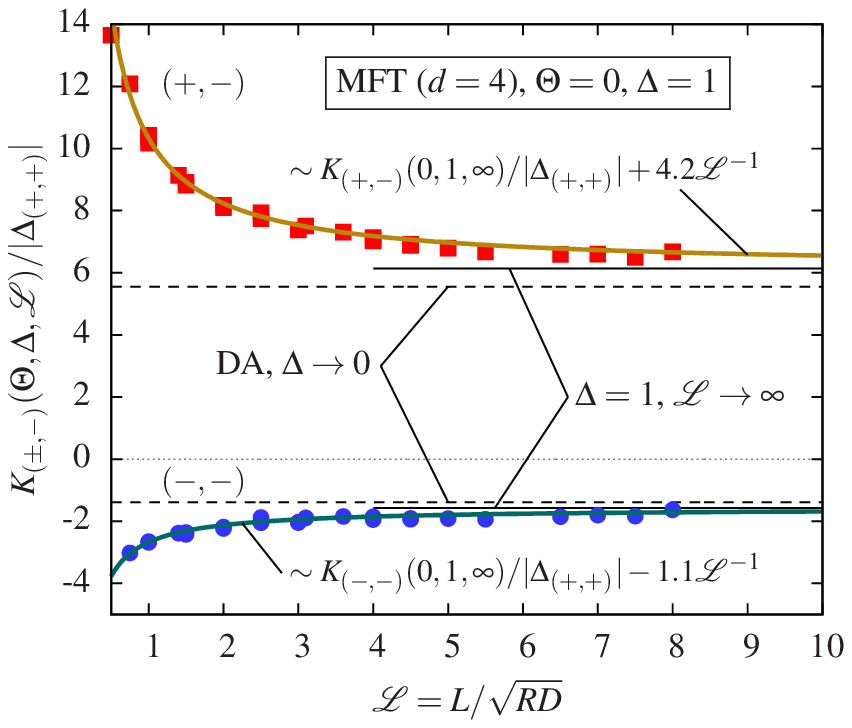}
  \else
  \includegraphics[width=11cm]{fig_05}
  \fi
\end{center}
  \clevercaption{%
  Reduced scaling functions $K_{(\pm,-)}(\Theta,\Delta,\Len)/|\Dpp|$ of the normal
  critical Casimir force for $(+,-)$ (top) and $(-,-)$ (bottom) BCs as a function
  of the rescaled length $\Len=L/\sqrt{RD}$ of the colloid.
  $K_{(\pm,-)}$ are shown for the specific choices $\Delta=1$ and $\Theta=0$ in
  order to test a limiting case with a view on the applicability of the DA.
  Indeed, for small values of $\Len$, the full MFT scaling functions significantly deviate
  from their asymptotic values $K_{(\pm,-)}(\Theta=0,\Delta=1,\Len\to\infty)/|\Dpp|$ for
  an infinitely extended cylinder shown as solid horizontal lines.
  However, for $\Len\gtrsim4$, they agree within $15\%$ with the limiting
  values, which are approached $\propto\Len^{-1}$
  as shown by the thick solid curves.
  Via least-square fits, for the present data we have obtained the universal MFT amplitudes
  $\kappa_{-}=-1.1\pm0.2$ and $\kappa_{+}=4.2\pm0.3$ (see \eref{eq:long}) describing 
  this asymptotic approach.
  The DA limits for $\Delta\to0$ (dashed horizontal lines) are independent of the value 
  of $\Len$ because within DA the effects of the cylinder ends are neglected.
  }   
  \label{fig:cyl-finite}
\end{figure}

In \fref{fig:cyl-finite} we show the reduced MFT scaling functions
$K_{(\pm,-)}(\Theta,\Delta,\Len)/|\Dpp|$ for the fixed values $\Theta=0$ and $\Delta=1$ as
a function of $\Len$.
The size of the symbols shown in \fref{fig:cyl-finite} indicates the estimated
numerical error, which is larger than for the limiting case $\Len\to\infty$.
Moreover, we show the corresponding values of the scaling functions $K_{(\pm,-)}$ for the limit
$\Len\to\infty$ and for $\Delta=1$ as solid horizontal lines.
The results obtained within the DA for $\Delta\to0$ [\eref{eq:cyl-force-da}] are
independent of the value of $\Len$ and are shown as dashed horizontal lines.

From \fref{fig:cyl-finite}, we can infer that for small values of $\Len\lesssim2$ the 
scaling function of the critical Casimir force for $\Delta=1$ deviates significantly 
from its corresponding expression obtained from the DA, as well as from its corresponding value attained
in the limit $\Len\to\infty$.
On the other hand, for $\Len\gtrsim4$, the deviation of the value of the critical Casimir force from its
limiting one for $\Len\to\infty$ is less than 15\% for both $(-,-)$ and $(+,-)$ BCs.
From \fref{fig:cyl-finite} we find that the numerically obtained scaling functions 
$K_{(\pm,-)}(\Theta,\Delta,\Len)$ are in accordance with the expected limiting behavior 
given in \eref{eq:long} already for $\Len\gtrsim1$.
From least-square fits we have obtained from our data for $\Theta=0$ and $\Delta=1$
the values $\kappa_{-}=-1.1\pm0.2$ and $\kappa_{+}=4.2\pm0.3$. In \fref{fig:cyl-finite}, the fits based on \eref{eq:long} are shown as thick solid curves and are in agreement with all data points within the estimated error bars.
According to the values found for $\kappa_\pm$, we note that, although the scaling functions
for $\Len\lesssim2$ clearly deviate from the DA values for $\Delta=1$, the 
ratio $\Kpm(\Theta=0,\Delta,\Len)/\Kmm(\Theta=0,\Delta,\Len)$ hardly changes as a function of $\Len$
and basically is in agreement with the DA ratio 
$\Kpm(\Theta=0,\Delta\to0,\Len)/\Kmm(\Theta=0,\Delta\to0,\Len)=-4$ in $d=4$.

Although we have analyzed the dependence of $K_{(\pm,-)}$ on $\Len$ only for a single value of $\Delta$, we expect this behavior to carry over to smaller values $\Delta\lesssim1$.
Moreover, since at $T=T_c$ (i.e., $\Theta=0$), the critical Casimir force exhibits its longest range, we
also expect the deviations of the scaling functions from their corresponding values 
for $\Len\to\infty$ to be less pronounced for nonzero values of $\Theta$, so that the investigated example
probes the strongest influence of the finite length of the cylinder with respect to the reliability of the DA.

From our analysis above we conclude that the DA provides a rather reliable approximation
for the scaling functions $K_{(\pm,-)}(\Theta,\Delta,\Len)$ for both $(\pm,-)$ BCs for a wide
range of parameters, which happen to be the relevant ones for experiments involving
colloids in binary liquid mixtures \cite{gambassi:2009,troendle:2010}.
In particular, the DA agrees rather well with the actual, full MFT data in the mixed state of the binary liquid at
$t>0$, for distance-to-radius ratios $\Delta\lesssim1/3$, as well as for elongated cylinders $\Len\gtrsim4$.
We expect these parameter ranges concerning the reliability of the DA to carry over accordingly to $d=3$.
Thus, in our subsequent analyses we make use of the DA in order to predict the critical Casimir interaction
between a cylindrical colloid and a chemically patterned substrate with even richer geometrical features.

\newpage
\section{Chemical step\label{sec:step}}
%
In this section we consider a cylindrical particle with  $(b)$ BC near a planar substrate patterned with
a single chemical step such that it exhibits $(a_<)$ BC for $x<0$ and $(a_>)$ BC for $x\ge0$ along the lateral 
direction $x$ (see \fref{fig:cyl_substrate}(c)).
The axis of the colloid is taken to be always parallel to the substrate surface.
The projection of the position of the center of the cylinder with respect to the $x$ direction
is denoted by $X$, so that for $X=0$ the center of the cylinder is located directly above the chemical step.
Whereas in Ref.~\cite{troendle:2010} the analysis of the critical Casimir force has been limited to an infinitely extended cylinder aligned parallel 
to the chemical step, i.e., perpendicular to the $x$ direction, here we consider rotations of the cylinder by an angle $\alpha$ 
with respect to the chemical step and discuss the dependence of the critical Casimir force on the cylinder length.
As a result of the dependence on the rotation angle $\alpha$, a critical Casimir \emph{torque} emerges.
For a chemically homogeneous colloid surface, due to the underlying symmetry $\alpha$ can be limited to the range $[0,\pi/2]$ without loss of generality.

\vspace*{-2em}
\subsection{Normal critical Casimir force and critical Casimir potential}
%
As an extension of the preceding section the normal critical Casimir force acting on a cylindrical colloid near a chemical \textit{s}tep takes the
following scaling form (compare \eref{eq:cyl-force-homog}):
\small
\begin{equation} 
  \label{eq:force-step}
  \Fstep(X,\alpha,D,R,L,T)=k_BT\frac{LR^{1/2}}{D^{d-1/2}}\Kstep(\Xi,\alpha,\Theta,\Delta,\Len),
\end{equation}
\normalsize
where $\Xi=X/\sqrt{RD}$ is the scaling variable corresponding to the lateral position of the cylinder.
Following the line of arguments in Ref.~\cite{troendle:2010} it is useful to introduce the following decomposition:
\small
\begin{multline} 
  \label{eq:force-step-split}
  \Kstep(\Xi,\alpha,\Theta,\Delta,\Len)=\frac{\Kasb(\Theta,\Delta,\Len)+\Kalb(\Theta,\Delta,\Len)}{2}\\
  +\frac{\Delta K(\Theta,\Delta,\Len)}{2}\psi_{(a_<|a_>,b)}(\Xi,\alpha,\Theta,\Delta,\Len),
\end{multline} 
\normalsize
where
\small
\begin{equation}
  \Delta K(\Theta,\Delta,\Len)\equiv\Kasb(\Theta,\Delta,\Len)-\Kalb(\Theta,\Delta,\Len)
  \label{eq:force-deltaK}
\end{equation}
\normalsize
is the difference between the scaling functions corresponding to the cases of the cylinder being located far away
either to the left ($<$) or to the right ($>$) from the chemical step.
The scaling function $\psi_{(a_<|a_>,b)}$ crosses over from $+1$ at $\Xi\to-\infty$ to $-1$ at $\Xi\to+\infty$ such
that the scaling functions of the laterally homogeneous substrates are recovered far from the step.
Accordingly, the critical Casimir potential $\Phistep(D)\equiv\int_D^\infty\upd D'\; \Fstep(D')$  
can be written as (compare \eref{eq:cyl-pot-homog})
\small
\begin{equation} 
  \label{eq:pot-step}
  \Phistep(X,\alpha,D,R,L,T)=k_BT\frac{LR^{1/2}}{D^{d-3/2}}\varthetastep(\Xi,\alpha,\Theta,\Delta,\Len)
\end{equation} 
\normalsize
with
\small
\begin{multline} 
  \label{eq:pot-step-split}
  \varthetastep(\Xi,\alpha,\Theta,\Delta,\Len)=\frac{\varthetaasb(\Theta,\Delta,\Len)+\varthetaalb(\Theta,\Delta,\Len)}{2}\\
  +\frac{\Delta\vartheta(\Theta,\Delta,\Len)}{2}\omega_{(a_<|a_>,b)}(\Xi,\alpha,\Theta,\Delta,\Len),
\end{multline}
\normalsize%
where
\small
\begin{equation}
  \label{eq:pot-delta-theta}
  \Delta \vartheta(\Theta,\Delta,\Len)\equiv\varthetaasb(\Theta,\Delta,\Len)-\varthetaalb(\Theta,\Delta,\Len)
\end{equation}
\normalsize
is the difference between the scaling functions of the laterally homogeneous substrates which are attained for $X\to\pm\infty$.
Thus, $\omega_{(a_<|a_>,b)}$ also crosses over from $+1$ at $\Xi\to-\infty$ to $-1$ at $\Xi\to+\infty$.

Similar to Eqs.~\eqref{eq:scalefunc_cyl_pot_relation} and \eqref{eq:scalefunc_cyl_force_relation}, valid for a homogeneous substrate, the scaling function \linebreak
$\varthetastep(\Xi,\alpha,\Theta,\Delta,\Len)$ of the potential in \eref{eq:pot-step} can be expressed in terms of 
the scaling function $\Kstep(\Xi, \alpha, \Theta,  \Delta,\Len)$ of the critical Casimir force\vspace*{-0.7em}
\small
\begin{equation}
\varthetastep(\Xi,\alpha,\Theta,\Delta,\Len) = \int\limits_1^\infty \upd z\ \frac{\Kstep(\Xi/\sqrt{z}, \alpha, z\, \Theta,  z\, \Delta,\Len)}{{z}^{d-1/2}}
\label{eq:scalefunc_cyl_pot_step_relation}
\vspace*{-0.7em}
\end{equation}
\normalsize
and vice versa
\small
\begin{align}
\label{eq:scalefunc_cyl_force_step_relation}
\Kstep(\Xi, \alpha, \Theta, \Delta,\Len) =
&(d-3/2) \varthetastep(\Xi,\alpha,\Theta,\Delta,\Len)\nonumber\\
&+ \frac{\Xi}{2}\,\frac{\partial}{\partial\Xi} \varthetastep(\Xi,\alpha,\Theta,\Delta,\Len)\nonumber\\
&- \Theta\,\frac{\partial}{\partial\Theta} \varthetastep(\Xi,\alpha,\Theta,\Delta,\Len)\nonumber\\
&- \Delta\,\frac{\partial}{\partial \Delta}\varthetastep(\Xi,\alpha,\Theta,\Delta,\Len).%
\end{align}
\normalsize

Within the limit $\Delta\to0$, the DA can be carried out semi-analytically.
Whereas this does not hold in general, within DA the scaling functions $\psi_{(a_<|a_>,b)}$ and $\omega_{(a_<|a_>,b)}$ are independent of the choice of $(a_<)$, $(a_>)$, and $(b)$ in the case of symmetry-breaking BCs $(\pm)$ and vanishing bulk field. 
Thus, the different choices of the BCs $(a_<|a_>,b)$ are reflected in the scaling function $\Kstep$ of the force [\eref{eq:force-step-split}] via $\Delta K$.
For $\{(+|-,-), (-|+,+)\}$ one has $\Delta K>0$ and for $\{(+|-,+), (-|+,-)\}$ one has $\Delta K<0$.
In Ref. \cite{troendle:2009} a comparison has been carried out between the DA and full MFT results for the case of a spherical colloid in $d=4$ opposite to a substrate with a chemical step. Although the full MFT results do not exhibit the exact antisymmetric shape of the DA results, it turns out that for $\Delta < 1/3$ they are in agreement with the DA on a quantitative level.
However, in order to indicate the general dependence on the BCs, in the remainder we keep the index $(a_<|a_>,b)$ of the scaling functions.

Within DA, the scaling functions $\psi_{(a_<|a_>,b)}$ and $\omega_{(a_<|a_>,b)}$ are, by construction, 
odd function with respect to the lateral position $\Xi$, so that 
$\psi_{(a_<|a_>,b)} (\Xi<0,\ldots)= -\psi_{(a_<|a_>,b)} (\lvert\Xi\rvert,\ldots)$ and 
$\omega_{(a_<|a_>,b)}(\Xi<0,\ldots)= -\omega_{(a_<|a_>,b)}(\lvert\Xi\rvert,\ldots)$.
It follows that $\psi_{(a_<|a_>,b)} (\Xi=0,\ldots)=\omega_{(a_<|a_>,b)} (\Xi=0,\ldots) = 0$. 
This implies that for the case of a particle positioned at $\Xi=0$, within DA,
\eref{eq:force-step-split} yields the scaling function $\Kstep$ of the force acting on the particle:\vspace*{-1em}
\ifTwocolumn
\small
\begin{multline}
\Kstep(\Xi=0,\alpha,\Theta,\Delta\to0,\Len)\\=
\frac{\Kasb(\Theta,\Delta\to0,\Len)+\Kalb(\Theta,\Delta\to0,\Len)}{2}.
\end{multline}
\normalsize
\else
\begin{equation}
\Kstep(\Xi=0,\alpha,\Theta,\Delta\to0,\Len)=\frac{\Kasb(\Theta,\Delta\to0,\Len)+\Kalb(\Theta,\Delta\to0,\Len)}{2}.
\end{equation}
\fi
Similarly \eref{eq:pot-step-split} yields the scaling function $\varthetastep$ of the potential:
\ifTwocolumn
\small
\begin{multline}
\varthetastep(\Xi=0,\alpha,\Theta,\Delta\to0,\Len)\\=
\frac{\varthetaasb(\Theta,\Delta\to0,\Len)+\varthetaalb(\Theta,\Delta\to0,\Len)}{2},
\end{multline}
\normalsize
\else
\begin{equation}
\varthetastep(\Xi=0,\alpha,\Theta,\Delta\to0,\Len)=\frac{\varthetaasb(\Theta,\Delta\to0,\Len)+\varthetaalb(\Theta,\Delta\to0,\Len)}{2},
\end{equation}
\fi
which is a consequence of the assumption of additivity of the forces underlying the DA.

The scaling function $\psi_{(a_<|a_>,b)}$ of the critical Casimir force within DA reads (see Appendix \ref{sec:app-da} and Ref. \cite{labbe:thesis} for an in-depth discussion of the derivation)
\small
\begin{align}
  \label{eq:step-force-da}
  \psi_{(a_<|a_>,b)}(\Xi>0,\alpha,\Theta,\Delta\to0,\Len) =
  &-|\epsilon| \psi_{(a_<|a_>,b)}^\parallel(\Xi_1,\Theta)\nonumber\\
  &+(\epsilon+1) \psi_{(a_<|a_>,b)}^\parallel(\Xi_2,\Theta)\nonumber\\
  &+\frac{2\cot|\alpha|}{\Len\Delta K(\Theta,\Delta,\Len)}\nonumber\\
  &\times\int_{1+\Xi_1^2/2}^{1+\Xi_2^2/2}\upd\beta\;\beta^{-d}\Delta k(\beta\Theta),
\end{align}
\normalsize
where $\Delta k=k_{(a_<,b)}-k_{(a_>,b)}$ is the difference between the scaling 
functions of the critical Casimir force acting on two planar walls with $\Asb$ and $\Alb$ BCs, respectively
[\eref{eq:planar-force}].
The quantities $\epsilon$, $\Xi_1$, and $\Xi_2$ are abbreviations for
\ifTwocolumn
\small
\begin{align}
  \epsilon &\equiv \frac{|\Xi|}{\Len\sin|\alpha|}-\frac{1}{2},\nonumber\\
\Xi_1 &\equiv \Len|\epsilon|\tan|\alpha|%
=\left(\text{sign}\,\epsilon\right)\left\{\frac{\lvert\Xi\rvert}{\cos\alpha}-\frac{\Len\sin\lvert\alpha\rvert}{2\cos\alpha}\right\}%
, \nonumber\\
\Xi_2 &\equiv \Len(\epsilon+1)\tan|\alpha|%
=\frac{\Xi}{\cos\alpha}-\frac{\Len\sin\lvert\alpha\rvert}{2\cos\alpha}%
.\label{eq:defs}
\end{align}
\normalsize
\else
\begin{equation}
  \label{eq:defs}
  \begin{gathered}
  \epsilon \equiv \frac{|\Xi|}{\Len\sin|\alpha|}-\frac{1}{2},\\
\Xi_1 \equiv \Len|\epsilon|\tan|\alpha|%
=\left(\text{sign}\,\epsilon\right)\left\{\frac{\lvert\Xi\rvert}{\cos\alpha}-\frac{\Len\sin\lvert\alpha\rvert}{2\cos\alpha}\right\}%
,\quad%
\Xi_2 \equiv \Len(\epsilon+1)\tan|\alpha|%
=\frac{\Xi}{\cos\alpha}-\frac{\Len\sin\lvert\alpha\rvert}{2\cos\alpha}%
.
\end{gathered}
\end{equation}
\fi

They describe the overlap between the projection of the cylinder 
axis onto the substrate and parts of the chemical step (see also \fref{fig:DA_cyl_sketch} in Appendix \ref{sec:app-da}).
In \eref{eq:step-force-da} $\psi_{(a_<|a_>,b)}^\parallel$ corresponds to the scaling function
for the critical Casimir force of a non-rotated cylinder ($\alpha=0$) with $(b)$ BC in front
of a chemical $(a_<|a_>)$ step which is given by \cite{troendle:2010} 
\ifTwocolumn
\small
\begin{multline} 
  \label{eq:psi-parallel}
  \psi_{(a_<|a_>,b)}^\parallel(\Xi
  >
  0,\Theta)=\\
    -1
    +
    \frac{\displaystyle\sqrt{2}
    \int_{1+\Xi^2/2}^\infty\upd\beta\,(\beta-1)^{-\frac{1}{2}}\,\beta^{-d}\Delta k(\Theta\beta)}
    {\Delta K(\Theta,\Delta\to0,\Len)}.
\end{multline}
\normalsize 
\else
\begin{equation}
  \label{eq:psi-parallel}
  \psi_{(a_<|a_>,b)}^\parallel(\Xi
  >
  0,\Theta)=\\
    -1
    +
    \frac{\sqrt{2}
    \int_{1+\Xi^2/2}^\infty\upd\beta\,(\beta-1)^{-\frac{1}{2}}\,\beta^{-d}\Delta k(\Theta\beta)}
    {\Delta K(\Theta,\Delta\to0,\Len)}.
\end{equation}
\fi
\par
Accordingly, the scaling function $\omega_{(a_<|a_>,b)}$ of the critical Casimir potential can be found by integrating over the critical Casimir force using Eqs.~\eqref{eq:cyl-force-da}, \eqref{eq:force-step-split}, \eqref{eq:force-deltaK}, \eqref{eq:scalefunc_cyl_pot_step_relation}, \eqref{eq:step-force-da}, and \eqref{eq:psi-parallel}, so that [compare \eref{eq:cyl-pot-da}]
\small
\begin{align} 
  \label{eq:step-pot-da}
  \omega_{(a_<|a_>,b)}(\Xi>0,\alpha,\Theta,\Delta\to0,\Len) =
  &-|\epsilon| \omega_{(a_<|a_>,b)}^\parallel(\Xi_1,\Theta)\nonumber\\
  &+(\epsilon+1) \omega_{(a_<|a_>,b)}^\parallel(\Xi_2,\Theta)\nonumber\\
  &+\frac{2\cot|\alpha|}{\Len\Delta\vartheta(\Theta,\Delta\to0,\Len)}\nonumber\\
  &\times I_\omega(\Xi_1,\Xi_2,\Theta),
\end{align} 
\normalsize
where
\small
\begin{multline} 
  \label{eq:omega-parallel}
  \omega_{(a_<|a_>,b)}^\parallel(\Xi
  >
  0,\Theta)=\\
    -1
    +
    \frac{\displaystyle\Xi^3\int_{1}^\infty\upd\beta\,(\sqrt{\beta}-1)(1+\beta\Xi^2/2)^{-d}\Delta k([1+\beta\Xi^2/2]\Theta)}
    {\Delta\vartheta(\Theta,\Delta\to0,\Len)},
\end{multline} 
\normalsize
is the scaling function for the critical Casimir potential of a non-rotated cylinder with $(b)$ BC in front of 
a $(a_<|a_>)$ step, and
\small
\begin{multline}
  \label{eq:step-pot-da-I}
  I_\omega(\Xi_1,\Xi_2,\Theta)=
  \int_{1+\Xi_1^2/2}^{1+\Xi_2^2/2}\upd\beta\;(\beta^{-d+1}-\beta^{-d})\Delta k(\beta\Theta)\\
  -\frac{\Xi_1^2}{2}\int_{1+\Xi_1^2/2}^\infty\upd\beta\frac{\Delta k(\beta\Theta)}{\beta^d}
  +\frac{\Xi_2^2}{2}\int_{1+\Xi_2^2/2}^\infty\upd\beta\frac{\Delta k(\beta\Theta)}{\beta^d}.
\end{multline}
\normalsize

Using partial integration, it turns out that $I_\omega$ can be expressed also as
\begin{equation}
I_\omega(\Xi_1,\Xi_2,\Theta)=\int_{1+\Xi_1^2/2}^{1+\Xi_2^2/2}\upd\beta\int_\beta^\infty\upd\gamma\;\gamma^{-d}\Delta k(\gamma\Theta).
\end{equation}
\par
Moreover, for $\alpha=0$ the scaling functions $\psi_{(a_<|a_>,b)}$ and $\omega_{(a_<|a_>,b)}$ reduce to the corresponding
expressions for a non-rotated cylinder with its axis parallel to the chemical step, i.e.,  
$\psi^\parallel_{(a_<|a_>,b)}$ and $\omega^\parallel_{(a_<|a_>,b)}$, respectively
\footnote{%
This can been seen explicitly by estimating the last integral in \eref{eq:step-force-da} and the integrals in \eref{eq:step-pot-da-I}
for $\alpha\to0$ via the mean value theorem of integration, using the relations 
$\Xi_1,\Xi_2\xrightarrow{\alpha\to0}\Xi$ and $\Xi_2^2-\Xi_1^2=2\Len\Xi\tan|\alpha|/\cos|\alpha|$. In \eref{eq:step-force-da}, this reduces the divergent prefactor $\cot\lvert\alpha\rvert$ to a constant. Similarly, in \eref{eq:step-force-da} the divergences in the limit $\alpha\to0$ of the terms $\propto \varepsilon$ are eliminated by the vanishing of the integrals in this limit, yielding the same constant as the last integral but with opposite sign, so that for $\alpha\to0$ all these terms cancel out except $\psi^\parallel_{(a_<|a_>,b)}$ and $\omega^\parallel_{(a_<|a_>,b)}$.}.
Thus, within DA, $\psi^\parallel_{(a_<|a_>,b)}$ and $\omega^\parallel_{(a_<|a_>,b)}$ are also odd functions of $\Xi$.
%
%
%
%
\par
For $\alpha=\pi/2$, i.e.,  the cylinder axis being aligned perpendicular to the chemical step, we find $\epsilon=\lvert\Xi\rvert/\Len-1/2$ and
\begin{multline}
  \label{eq:alpha90}
  \psi_{(a_<|a_>,b)}(\Xi\gtrless0,\alpha=\tfrac{\pi}{2},\Theta,\Delta\to0,\Len)=\\
  \omega_{(a_<|a_>,b)}(\Xi\gtrless0,\alpha=\tfrac{\pi}{2},\Theta,\Delta\to0,\Len)=\mp 1\mp E(\epsilon)
  \quad\text{with }\\ E(\epsilon) = \left\{
  \begin{split}
    0,\;\;\; & \epsilon\ge0  \\
    2\epsilon,\;\;\; & \epsilon<0 \\
  \end{split}\right\}\hspace{5em}
\end{multline}
and $E(\epsilon)\xrightarrow{\Xi\to\pm\infty}0$, which follows from Eqs.~\eqref{eq:step-force-da} - \eqref{eq:psi-parallel} and Eqs.~\eqref{eq:step-pot-da} - \eqref{eq:step-pot-da-I}, respectively.

Thus, for $\alpha=\pi/2$ the scaling functions for both the normal critical Casimir force and for
the corresponding potential interpolate as a function of $\Xi$ linearly between the two scaling functions of the laterally homogeneous substrates. Within DA, these two scaling function become valid for $X\ge L/2$ and $X\le -L/2$, respectively.

Note that although, according to \eref{eq:alpha90}, for $\alpha=\pi/2$ $\psi_{(a_<|a_>,b)}$ and $\omega_{(a_<|a_>,b)}$ are identical,
they contribute to the corresponding scaling functions of the critical Casimir force [\eref{eq:force-step-split}]
and potential [\eref{eq:pot-step-split}] with different prefactors $\Delta K(\Theta,\Delta\to 0,\Len)/2$ and 
$\Delta\vartheta(\Theta,\Delta\to 0,\Len)/2$, respectively.
In Appendices \ref{sec:app-crit} and \ref{sec:app-far} we provide explicit expressions for the scaling functions
$\psi_{(a_<|a_>,b)}$ and $\omega_{(a_<|a_>,b)}$ for the special cases
$\Theta=0$ and $\Theta\gg1$, respectively, for which the corresponding integrations in Eqs.~\eqref{eq:step-force-da} and \eqref{eq:step-pot-da}
can be carried out analytically.

\begin{figure}[!htbp]
    \centering
   \ifTwocolumn
  \includegraphics[width=7.5cm]{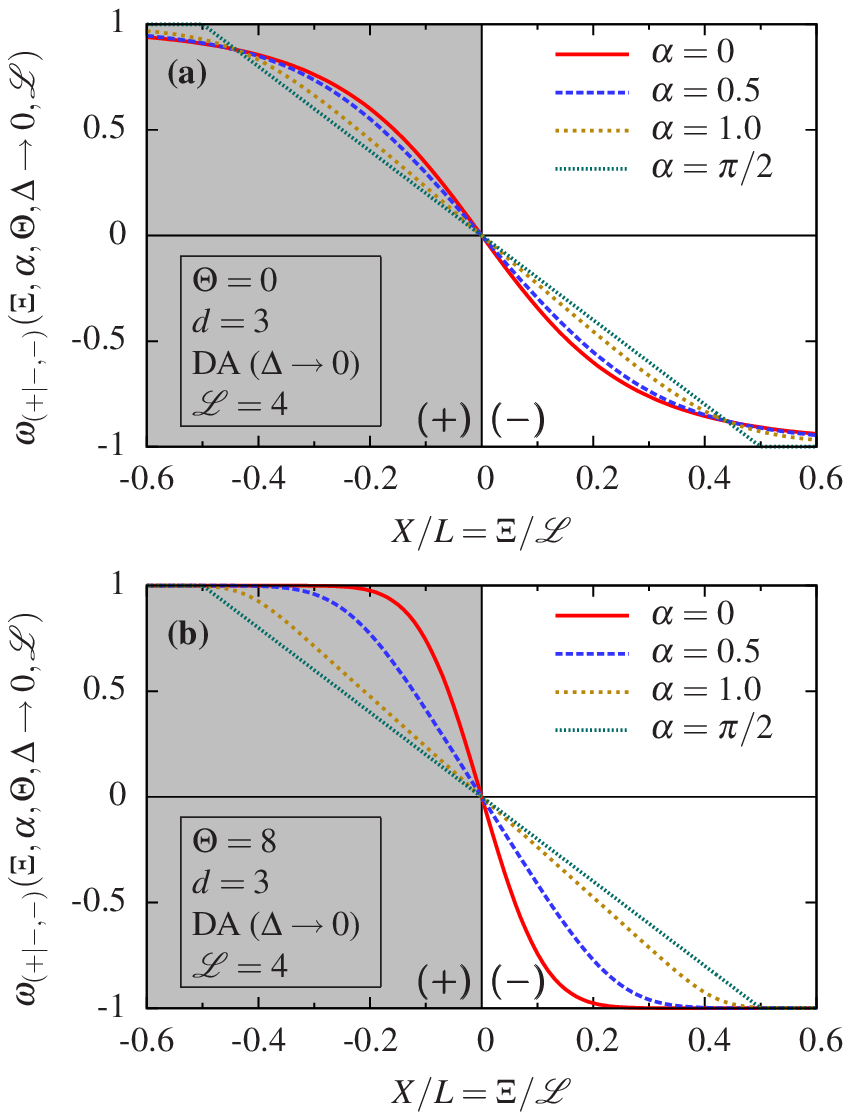}
  \else
  \includegraphics[width=11cm]{fig_06}
  \fi
    \clevercaption{Scaling function $\omega_{(a_<|a_>,b)}$ [Eqs.~\eqref{eq:pot-step} -- \eqref{eq:pot-delta-theta} and \eqref{eq:step-pot-da} -- \eqref{eq:step-pot-da-I}]
    of the critical Casimir potential as obtained within the DA for 
    $d=3$ as a function of the lateral particle position $X$ in units of the length $L$
    of the cylindrical colloidal particle.
    The scaling function $\omega_{(a_<|a_>,b)}$ is shown for four angles $\alpha$, a fixed value $\Len=4$,
    and $\Theta=0$ in (a), as well as $\Theta=8$ in (b). 
    The BCs are chosen such that $(a_<)=(+)$, $(a_>)=(-)$, and $(b)=(-)$, i.e., 
    we specifically depict $\omega_{(+|-,-)}$, which within the DA is identical to the other possible choices of $(\pm)$ BCs.
    For $\alpha=\pi/2$ the potential linearly interpolates between its homogeneous counterparts and is equal to them
    for $|X|/L>0.5$ [\eref{eq:alpha90}], whereas for non-perpendicular orientations of the cylinder relative to the chemical step the potential is smoothened.
    The expressions found within DA depend on the scaling functions $\Delta k=k_{(a_<,b)}-k_{(a_>,b)}$ of the force for the film geometry, except for $\Theta=0$. For all plots in $d=3$, the input of these film scaling functions is taken from the Monte Carlo estimates referred to as ``approximation (\textit{i})'' in Figs.~9 and 10 of Ref.~\cite{vasilyev:2009}.
    We note that the corresponding analytic approximations for $\Theta\gg 1$ given in \eref{eq:erf2} in Appendix \ref{sec:app-far}
    fall on top of the data shown for $\Theta=8$ in (b).
    }
    \label{fig:cyltilted_omega}
\end{figure}
The scaling function $\omega_{(a_<|a_>,b)}$ of the potential as obtained within the DA in $d=3$ is shown in 
\fref{fig:cyltilted_omega}(a) for $\Theta=0$, i.e., at the bulk critical point 
and in Fig. 6(b) for $\Theta=8$ corresponding to a higher temperature.
Within the DA, a cylindrical particle oriented perpendicular to the chemical step ($\alpha=\pi/2$) interacts
with the chemical step only for positions $|X|<L/2$. 
The overlap of the particle projection with the chemical step varies linearly with the position X, which leads as a function of $\Xi$
to a linear interpolation between $+1$ and $-1$ of the scaling function $\omega_{(a_<|a_>,b)}$ [\eref{eq:alpha90}]. 
Due to the DA, for $\alpha=\pi/2$ there is a non-analyticity at $\lvert X \rvert=L/2$.
For orientations $\alpha < \pi/2$, the interaction potential is smoothened due to the curved surface of the cylinder.
As can be inferred from \fref{fig:cyltilted_omega}(a), for $\Theta=0$ as a function of the colloid position the transition from the repulsive
to the attractive side of the substrate is rather smooth for all
orientations $\alpha$ of the cylinder (see \eref{eq:omega-0} in Appendix \ref{sec:app-crit}).
On the other hand, for the case $\Theta=8$ shown in \fref{fig:cyltilted_omega}(b) the system is far away from criticality
so that the force for the film geometry decays exponentially [\eref{eq:exponential-decay}] and
the crossover of $\omega_{(a_<|a_>,b)}$ between $+1$ and $-1$ for $\alpha<\pi/2$ is steeper as compared with the one seen in
\fref{fig:cyltilted_omega}(a).
From our analysis for $(\pm,-)$ BCs we find that for 
$\Theta\gtrsim4$ the numerically obtained scaling functions based on Monte-Carlo data for the film geometry \cite{vasilyev:2009} are in good agreement with the corresponding approximation for $\Theta\gg1$ given in 
\eref{eq:erf2} in Appendix \ref{sec:app-far}.

\subsection{Critical Casimir torque \label{sec:torque}}

Since the critical Casimir potential depends on the angle $\alpha$ between the axis of the cylinder 
and the chemical \textit{s}tep, a critical Casimir torque $\tau_s$ acting on the particle arises.
The torque is a vector in the direction of the substrate normal with $\tau_s=\frac{d}{d\alpha}\Phi_s$ as the only nonzero component.
The orientation of the particle being parallel to the chemical step corresponds to $\alpha=0^\circ$, 
while an orthogonal orientation corresponds to $\alpha=90^\circ$, 
so that a positive torque $\tau_s$, i.e., an increase of $\Phi_s$ upon an increase of $\alpha$, leads to a preferred parallel alignment and negative torques to the preference of the perpendicular orientation.

\begin{figure}
    \centering
   \ifTwocolumn
  \includegraphics[width=8cm]{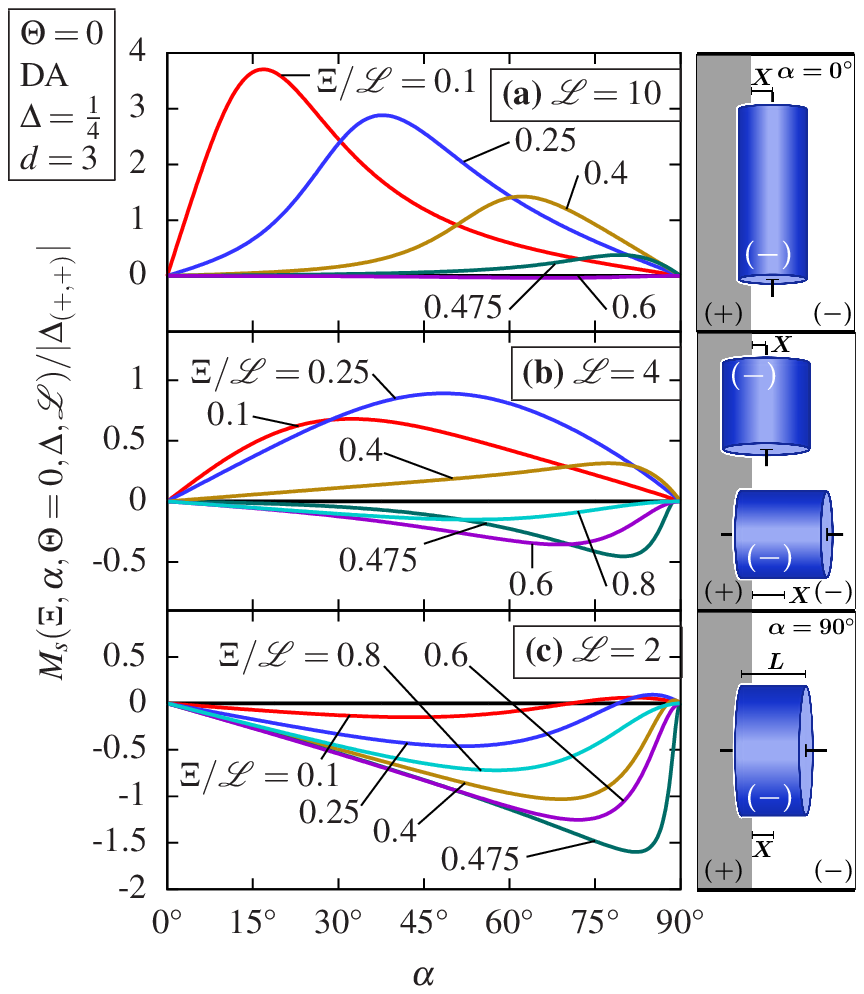}
  \else
  \includegraphics[width=11cm]{fig_07}
  \fi
    \clevercaption{Scaling function $M_s(\Xi,\alpha, \Theta=0,\Delta\to0,\Len)$ for the torque in 
    $d=3$ spatial dimensions as obtained within the DA [\eref{eq:cyl_tilted_Ms}] as a function of the orientation 
    $\alpha$ of the colloid relative to a chemical step [\fref{fig:cyl_substrate}(c)]. 
    The suitably normalized scaling function is shown for three different aspect ratios of the cylindrical colloids,
    i.e., for $\Len=L/\sqrt{RD}=10$ in (a), for $\Len=4$ in (b), and for $\Len=2$ in (c) as well as for various lateral distances 
    $\Xi/\Len=X/L>0$ from the chemical step.
    For negative values of $\Xi$, the scaling function can be obtained via a point reflection, i.e., 
    $M_s(\Xi<0,\cdots)=-M_s(-\Xi,\cdots)$.
    The solvent is considered to be at its bulk critical point $\Theta=0$. 
    For rod-like particles as in (a), we find for $0<X/L=\Xi/\Len<0.5$ the torque to be always positive,
    which leads to a preferred alignment parallel to the chemical step, as sketched right next to the graph.
    On the other hand for disk-like particles, as for $\Len = 2$ in (c), the torque is negative 
    for positive values of $X/L$,
    so that the colloid self-aligns perpendicular to the chemical step, as indicated in the sketch next to the graph.
    For the intermediate case $\Len=4$ in (b) we find 
    both negative and positive values of the critical Casimir torque for $X/L>0$, depending on the lateral
    position of the colloid.
    (The sketches next to the graphs correspond to aspect ratios $L/R=\Len\sqrt{\Delta}$ obtained for $\Delta=D/R=0.25$.)
    }
    \label{fig:cyltilted_torque}
\end{figure}

Based on \eref{eq:pot-step}, the critical
Casimir torque acting on the cylindrical particle can be written in the
following scaling form:
\small
\begin{equation}
  \tau_s(X,\alpha, D, R, L, T) = k_B T\, \frac{L R^{1/2}}{D^{d-3/2}}\,M_s(\Xi,\alpha, \Theta,\Delta,\Len),
\end{equation}
\normalsize
where the scaling function $M_s$ is given by [Eqs.~\eqref{eq:pot-step-split} and \eqref{eq:pot-delta-theta}]
\small
\begin{equation}
  \label{eq:msdef}
  M_s(\Xi,\alpha, \Theta,\Delta,\Len) =
  \frac{\Delta\vartheta(\Theta,\Delta,\Len)}{2} \frac{d}{d\alpha} \omega_{(a_<|a_>,b)}(\Xi,\alpha, \Theta,\Delta, \Len).
\end{equation}
\normalsize
In Appendix \ref{sec:app-torque} we determine $M_s$ for $\Delta\to 0$ within DA:
\small
\begin{multline}
  M_s(\Xi\gtrless 0,\alpha, \Theta,\Delta\to0,\Len) =
\mp \frac{1}{\sin^2\alpha}\\ \Bigg\{ \frac{1}{\Len}\,I_\omega(\Xi_1, \Xi_2,\Theta)
   + \frac{\Delta\vartheta(\Theta,\Delta\to0,\Len)}{2}
   \frac{\lvert\Xi\rvert}{\Len}\cos\alpha\\\times
   \left[ \omega^\parallel_{(a_<|a_>,b)}(\Xi_2,\Theta)
          -\sgn(\epsilon)\omega^\parallel_{(a_<|a_>,b)}(\Xi_1,\Theta)
          \right]\Bigg\}.
\label{eq:cyl_tilted_Ms}
\end{multline}
\normalsize
From \eref{eq:msdef} we infer that the critical Casimir torque is also an odd function of 
$\Xi$ and the signs in \eref{eq:cyl_tilted_Ms} are determined by $\sgn(\Xi)$ following from the definition 
of $\omega_{(a_<|a_>,b)}$ in \eref{eq:pot-step-split}.
Note that in general $I_\omega$, $\Delta\vartheta$, and $\omega^\parallel$ depend on the choice of BCs $(a_<)$, $(a_>)$, and $(b)$, whereas within DA $\omega^\parallel$ does not.
The dependence on the relative position $X/L=\Xi/\Len$ is illustrated in \fref{fig:cyltilted_torque} where we present the scaling function $M_s$ as 
obtained within the DA for $d=3$ as a function of the rotation angle $\alpha$ with the temperature fixed at its critical value, i.e., $\Theta=0$.
The relative position $X/L=\Xi/\Len$ is independent of the aspect ratio $L/R=\Len\sqrt{\Delta}$ of the particle; therefore the shape of the 
particle affects the torque only through the scaling variable $\Len=L/\sqrt{R D}$. 
For negative values of $\Xi$, the scaling function can be obtained via a point reflection, i.e., 
$M_s(\Xi<0,\cdots)=-M_s(-\Xi,\cdots)$.

Our results show that for large aspect ratios $L/R$ (i.e., rod-like particles), for which $\Len \gg 1$,
the torque acting on the colloid is positive for $0<X/L=\Xi/\Len<0.5$ and basically vanishes for $X/L=\Xi/\Len>0.5$ for all
rotation angles $\alpha\in[0,\pi/2]$. 
As can be seen in \fref{fig:cyltilted_torque}(a) for a particle with $\Len=10$, the torque vanishes when the particle 
is orientated parallel ($\alpha=0^\circ$) or perpendicular ($\alpha=90^\circ$) relative to the chemical step on the substrate. 
For $X/L>0$, the torque is positive and reaches a maximum value at an intermediate angle, so that the orientation with $\alpha=0$ is stable against rotations of the particle, 
whereas the perpendicular orientation is unstable and thus the rod-like particles with $\Len=10$ prefer to orientate themselves parallel to the chemical step.
For $X/L<0$, due to its above mentioned antisymmetry, the torque is negative, so that in this case the orientation with $\alpha=90^\circ$ is stable against rotations of the particle,
whereas the parallel orientation becomes unstable, in contrast to the case $X/L>0$.
As shown in \fref{fig:cyltilted_torque}(b), for smaller aspect ratios $L/R$ and $\Len\simeq4$ the torque changes sign upon varying the position $\Xi/\Len$ of the colloid.
The torque is positive if the particle is close to the step and the maximal strength of the torque first increases with the relative position $\Xi/\Len$, 
but then decreases and finally the torque changes into the opposite direction. 
This sensitivity of the orientation with respect to the geometrical features is due to the comparable length scales of the particle length $L$ and the radius $R$. 
For disk-like particles with $\Len \lesssim 2$ as shown in \fref{fig:cyltilted_torque}(c), we find that for $X/L>0$ the torque is mostly negative for all orientations of the 
particle, so that in this case the perpendicular orientation is the preferred one, whereas for $X/L<0$ the torque is positive and the parallel colloidal orientation is the preferred one.
Our results obtained within the DA for $\Len=2$ indicate a change of sign of the critical Casimir torque at angles $\alpha=\alpha_0\gtrsim70^\circ$ for $0.1<\Xi/\Len<0.5$. However,
for larger values of $\alpha>\alpha_0$ the magnitude of the scaling function $M_s$ is very small compared with the Casimir amplitude $|\Dpp|$.

\begin{figure}
    \ifTwocolumn \else\vspace*{-2cm}\fi
    \centering
   \ifTwocolumn
  \includegraphics[width=8cm]{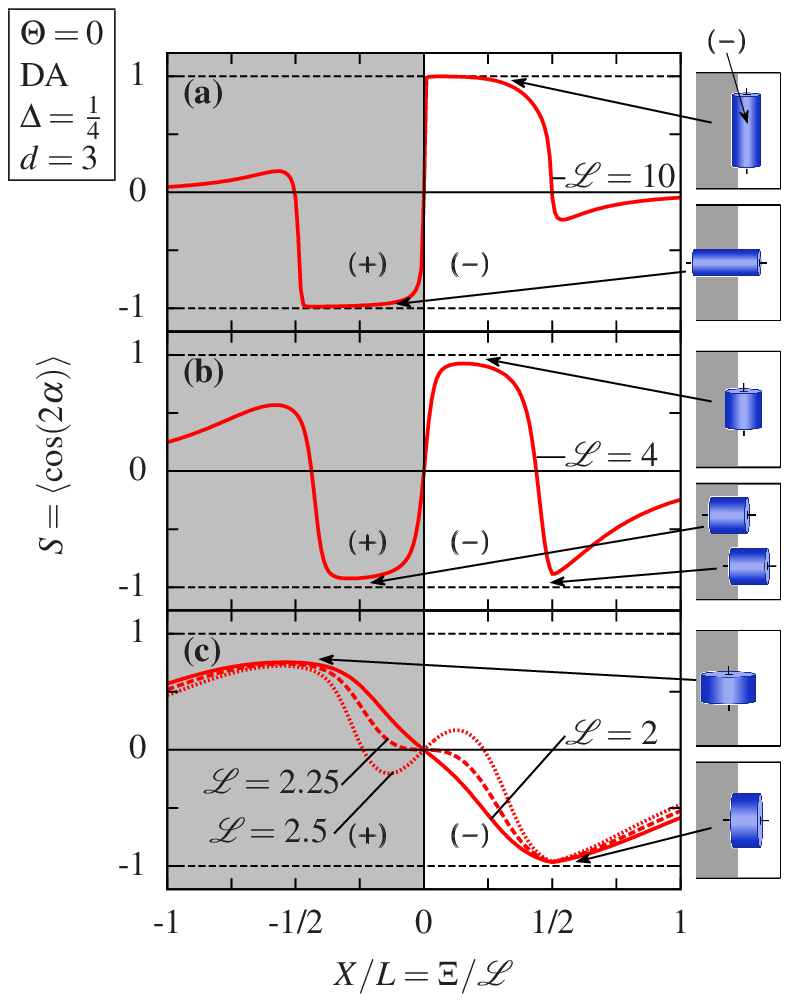}
  \else
  \includegraphics[width=11cm]{fig_08}
  \fi
    \clevercaption{%
    Planar nematic order parameter $S$ [\eref{eq:orderpara}] for a cylindrical colloid close to 
    a chemical step as obtained within the DA for $d=3$ and $\Delta=D/R=1/4$ as a function
    of the lateral particle position $X$ in units of its length $L$. The solvent is taken to be at its bulk critical point $\Theta = 0$.
    For rod-like particles with $\Len=10$ in (a), upon increasing $X$ the nematic order parameter $S$ changes abruptly from
    $S\simeq-1$ to $S\simeq+1$ at $\Xi=0$, corresponding to a change from a preferred colloid orientation perpendicular to the step
    to an orientation parallel to the step. Both configurations are very stable against thermal fluctuations.
    At $|\Xi/\Len|\simeq0.5$ in (a) $S$ again changes sign but it attains only small values
    for $|\Xi/\Len|>0.5$, corresponding to a weak preference of the colloid orientation, and tends
    to a uniform angular distribution ($S=0$) for $|\Xi/\Len|\gtrsim1$.
    For a shorter cylinder with $\Len=4$, in (b) the behavior for small values of $|\Xi/\Len|$ is similar
    as in (a), but the alignment at $|\Xi/\Len|\simeq0.5$ becomes very pronounced; $S$ is close
    to $-1$ for $\Xi/\Len=0.5$ which corresponds to a strong orientational alignment of the cylinder perpendicular to the chemical step.
    For disk-like particles with $\Len=2$, $\Len=2.25$, and $\Len=2.5$ the behavior of $S$ is different.
    Upon lowering $\Len$ the order parameter extrema close to $\Xi/\Len=0$ disappear and the 
    angular distribution becomes almost uniform (i.e., $|S|$ is small).
    On the other hand the alignment at $|\Xi/\Len|\simeq0.5$ is pronounced, but with the 
    opposite preference of the orientations as compared with the case $\Len=10$ and $\lvert\Xi/\Len\rvert<0.5$ in (a).
    The sketches next to the graphs, correspond to aspect ratios $L/R=\Len\sqrt{\Delta}=\Len/2$.
    }
    \label{fig:cyltilted_orderpara}
\end{figure}
In order to analyze the rotational orientation of the cylinder and its statistical characteristics with respect to thermal fluctuations
in more detail, we investigate the planar nematic order parameter $S$ defined as \cite{Straley:1971,Frenkel:1985}
\small
\begin{equation} 
  \label{eq:orderpara}
  S\equiv\langle\cos(2\alpha)\rangle=\frac{1}{N}\int\limits_0^{\pi/2}\upd\alpha\cos(2\alpha)\exp\left\{-\frac{\Phistep(X,\alpha,D,R,L,T)}{k_BT}\right\},
\end{equation}
\normalsize
where the normalization constant is given by $N=\int_0^{\pi/2}\upd\alpha\exp\left\{-{\Phistep(X,\alpha,D,R,L,T)}/{k_BT}\right\}$.
$S=1$ corresponds to perfect alignment of the cylindrical colloid parallel to the chemical step ($\alpha=0$), whereas $S=-1$ corresponds to perfect
alignment perpendicular to the step ($\alpha=90^\circ$).
Isotropic orientation is characterized by $S=0$.
In \fref{fig:cyltilted_orderpara} the nematic order parameter $S$ as obtained within the DA for $\Xi=0$ and $\Delta=1/4$ is shown for the same 
values of $\Len$ as in \fref{fig:cyltilted_torque} as a function of the relative position $X/L=\Xi/\Len$.
As can be inferred from \fref{fig:cyltilted_orderpara}(a), a rod-like particle with $\Len=10$ exhibits a strong rotational alignment when its center is close
to the chemical step. Whereas for $-0.5\lesssim\Xi/\Len<0$ the cylinder is strongly aligned perpendicular to the step due to the critical Casimir
torque, it abruptly changes orientation upon crossing the chemical step at $\Xi=0$.
For $0<\Xi/\Len\lesssim0.5$ the cylinder is aligned parallel to the step, exploiting fully the attractive critical Casimir interaction between 
surfaces of same chemical preference.
At $|\Xi/\Len|\simeq0.5$ the nematic order parameter $S$ again changes its sign. 
However, for $0.5 \lesssim |\Xi/\Len| \lesssim 1$ the magnitude of $S$ is rather small and vanishes for $|\Xi/\Len|\simeq1$, 
corresponding to a uniform angular distribution.

For a reduced cylinder length $\Len=4$, the change of the sign of $S$ at $\lvert\Xi/\Len\rvert\approx0.5$ becomes much more pronounced [see \fref{fig:cyltilted_orderpara}(b)].
Whereas close to the chemical step at $\Xi=0$ the behavior of the order parameter $S$ resembles the one for $\Len=10$ in \fref{fig:cyltilted_orderpara}(a) (but less abruptly),
a strong orientational alignment of the cylinder perpendicular to the step ($S=-1$) develops at $\Xi/\Len\simeq0.5$.
In addition for $\Xi/\Len\lesssim -0.5$ the degree of orientational order is higher than the corresponding one of the rod-like particle with $\Len=10$ in 
\fref{fig:cyltilted_orderpara}(a). Thus, as a function of its lateral position a cylindrical particle of reduced length $\Len=4$ exhibits 
various changes of its preferred orientation parallel or perpendicular to the chemical step.

For even smaller values of $\Len$, i.e., disk-like particles, the strong orientational alignment close to $\Xi/\Len=0$ disappears in that the nematic order parameter $S$ aquires a small amplitude, as can be inferred from  \fref{fig:cyltilted_orderpara}(c).
In addition, $S$ flips upon lowering $\Len$, such that for $\Len=2$ the particles align with their axis
parallel to the step for $\Xi<0$ and perpendicular to it for $\Xi>0$.
Moreover, the change between these two orientations as function of $\Xi$ is much smoother as compared with the case of rod-like particles in \fref{fig:cyltilted_orderpara}(a).
This is due to the relatively small strength of the critical Casimir torque for small values of $\Xi/\Len$, as shown in \fref{fig:cyltilted_torque}(c).
A change of sign of $M_s(\alpha)$ and the accompanying reversal of stability of the corresponding configurations signal the presence of competing minima in the free energy landscape. For \fref{fig:cyltilted_orderpara}(c) those are very shallow in units of $k_B T$ and therefore easily washed out by thermal fluctuations.

\begin{figure}
    \centering
   \ifTwocolumn
  \includegraphics[width=8cm]{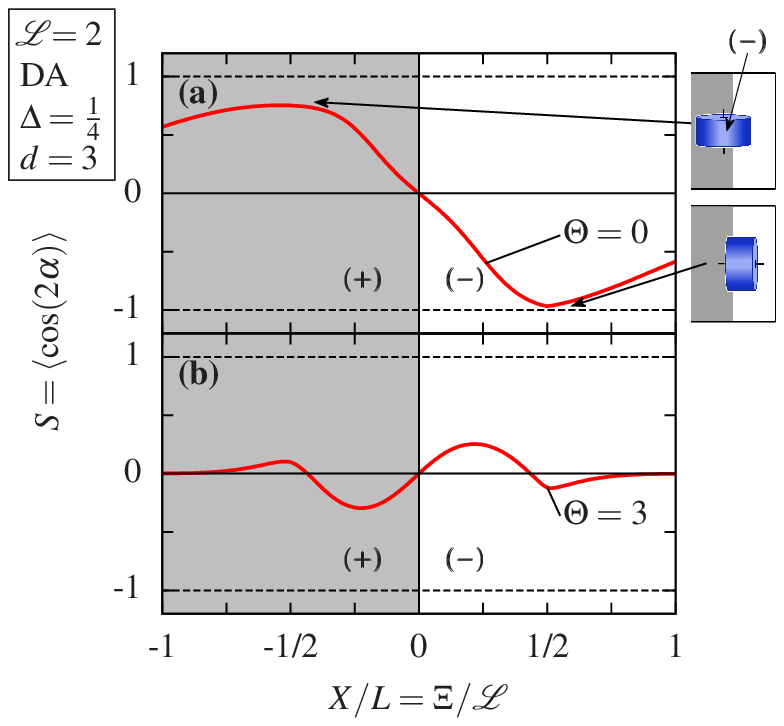}
  \else
  \includegraphics[width=11cm]{fig_09}
  \fi
    \clevercaption{%
    Planar nematic order parameter $S$ [\eref{eq:orderpara}] for a cylindrical colloid close to 
    a chemical step as obtained within the DA for $d=3$, $\Delta=D/R=1/4$, and $\Len = 2$ as a function
    of the lateral particle position $X$ in units of its length $L$. For such a disk-like particle we find a
    preferred parallel [perpendicular] orientation of the particle axis relative to the chemical step for $\Xi/\Len < 0$ [$\Xi/\Len > 0$]
    at the bulk critical point ($\Theta=0$) in (a) [same as in \fref{fig:cyltilted_orderpara}(c)]. On the other hand, for 
    a high temperature corresponding to $\Theta=3$ in (b), the nematic order parameter changes sign at $\Xi=0$ 
    and $|\Xi/\Len|\approx 0.4$. Thus, for $\lvert\Xi/\Len\rvert\lesssim0.4$ the preferential orientation of the disk-like particle switches upon varying temperature.
}
    \label{fig:cyltilted_orderpara_L2}
\end{figure}
For a disk-like particle with a reduced length $\Len=2$ %
\fref{fig:cyltilted_orderpara_L2} illustrates the temperature dependence of the orientational order parameter profile $S(X/L)$ by comparing the system at bulk criticality $\Theta=0$ [\fref{fig:cyltilted_orderpara_L2}(a)] and off criticality $\Theta=3$ [\fref{fig:cyltilted_orderpara_L2}(b)].
As discussed above in \fref{fig:cyltilted_orderpara}(c), in \fref{fig:cyltilted_orderpara_L2}(a) for $\Theta=0$ 
the critical Casimir torque leads to a preferential alignment of the cylinder axis parallel to the chemical step for $\Xi<0$ 
and perpendicular to the step for $\Xi>0$. However, for $\Theta=3$ in \fref{fig:cyltilted_orderpara_L2}(b) the nematic order 
parameter $S$ changes sign
at $\Xi=0$ and at $|\Xi/\Len|\approx 0.4$. Thus, whenever the perpendicular [parallel] orientation is the preferred one 
at the bulk critical point for $|\Xi/\Len|\lesssim 0.4$ as sketched in \fref{fig:cyltilted_orderpara_L2}(a), 
the disk-like colloid prefers a parallel [perpendicular] orientation at higher temperatures as sketched in \fref{fig:cyltilted_orderpara_L2}(b).
Accordingly, the orientation of a disk-like colloid near a chemical step can be reversibly and continuously switched by minute temperature changes.
We attribute this behavior to the fact that the ratio of the strengths of the critical Casimir forces in the film geometry for $(+,-)$ and
$(-,-)$ BCs varies as function of $\Theta$.
Whereas close to $T_c$ the critical Casimir force for $(+,-)$ BCs is much stronger than for $(-,-)$ BCs,
both become comparable in strength for $\Theta\gg1$ (see also the different scales in Figs.~\ref{fig:homog_pp} 
and \ref{fig:homog_pm}).
However, the maximal absolute value of the nematic order parameter $S$ for $\Theta=3$ in \fref{fig:cyltilted_orderpara_L2}(b) is rather
small so that the degree of orientational order is low. Upon increasing $\Theta$ the nematic order parameter $S$ vanishes gradually
and the angular distribution of the colloids becomes uniform.

We note that within the DA the range of the effective interaction of the colloid with the substrate along the direction normal to the cylinder axis tends
to be overestimated due to the parabolic distance approximation in \eref{eq:parabolic}. 
However, this is less important far away from criticality because the scaling function of this potential decays 
exponentially with respect to the surface-to-surface distance between the particle and the substrate, and within DA
contributions of surface elements at the ends of the cylinder become negligible.
On the other hand, within DA we expect the torque to be underestimated in the regime of disk-like particles, 
similar to the normal critical Casimir force as discussed above. 
However, the ratio of the forces $\Kpm(\Theta=0,\Delta,\Len)/\Kmm(\Theta=0,\Delta,\Len)$ is maintained constant even for small values of $\Len\sim1$ (see \fref{fig:cyl-finite}). 
Thus, we expect that these deficiencies of the DA do not affect the sign of the torque and the qualitative results for the particle orientation presented
above for $\Len\ge2$, concerning the distinct behavior of rod-like and disk-like particles.

\vspace{-1.5em}
\section{Janus cylinder close to a periodically striped substrate\label{sec:compass}}
\begin{figure}[ht!]
\ifTwocolumn \else\vspace*{-2cm}\fi
  \centering
   \ifTwocolumn
  \includegraphics[width=7.5cm]{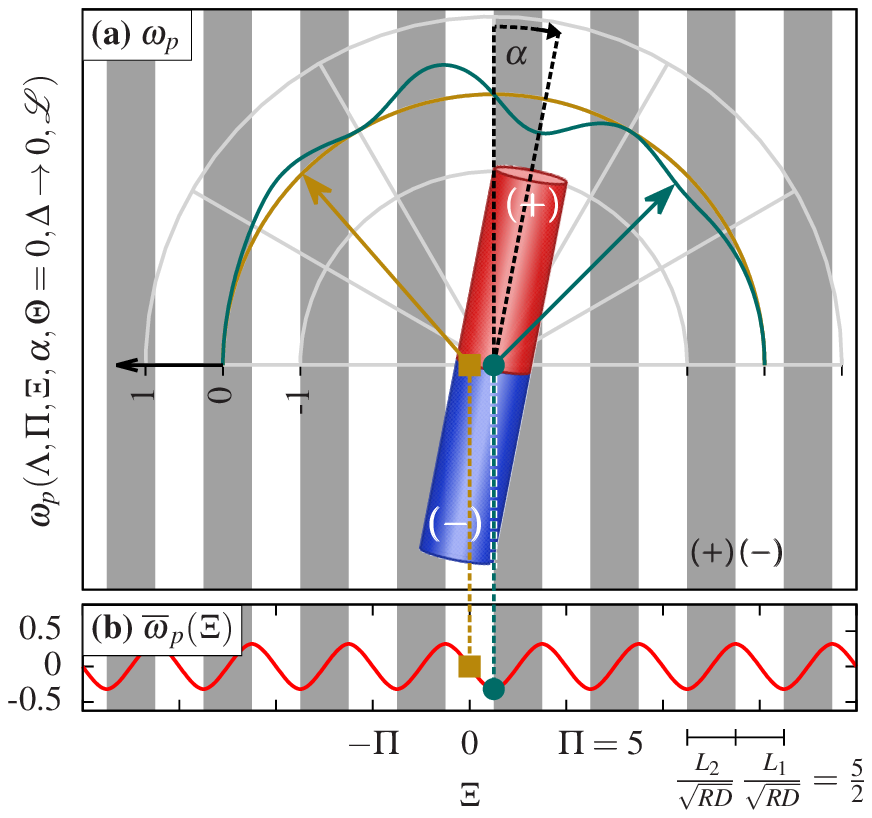}
  \else
  \vspace{-0.5em}\includegraphics[width=11cm]{fig_10}
  \fi
  \clevercaption{%
  Scaling function of the critical Casimir potential $\omega_p(\Lambda, \Pi, \Xi, \alpha, \Theta=0, \Delta\to 0, \Len)$ 
  [\eref{eq:omega_p}] within DA for $d=3$ and $\Theta=0$ acting on a Janus cylinder with opposite $(+)$ and $(-)$ BCs (red and blue areas, respectively)
  and of reduced length $\Len=L/\sqrt{RD}=20$  near a chemically patterned substrate.
  The substrate is periodically patterned with $(a_1)=(-)$ BC on one kind of stripes (white areas) and $(a_2)=(+)$ BC
  on the other kind of stripes (shaded areas).
  Due to this choice of the BCs, the red (blue) part of the Janus cylinder is attracted by the shaded (white)
  stripes and repelled by the others.
  The geometry of the pattern is characterized by $\Pi=P/\sqrt{RD}=5$ and $\Lambda=L_1/\sqrt{RD}=5/2$,
  where $L_1=L_2=P/2=L/8$ is the width of the stripes.
  The Janus cylinder, located at lateral position $X$ (so that $X=0$ corresponds to the center of a stripe with $(-)$ BC),
  is rotated by an angle $\alpha\in[-\pi/2,\pi/2]$ relative to the translationally invariant direction of the stripes.
  The normalized scaling function $\omega_p\in[-1,1]$ is shown in (a) as a function of $\alpha$ for the two
  colloid positions $X=0$ ($\Xi=0$, yellow curve) and $X=P/4$ ($\Xi=5/4$, green curve and illustrated by the sketched cylinder).
  The greyish curves are semi-circles around the green dot. Accordingly, for each point on the green line one can infer the corresponding 
  angle $\alpha$ by drawing the green arrow footed at $X\neq 0$. Consequently, the angles belonging to points on the yellow line can be 
  read off from the yellow arrow which, however, is footed at $X=0$. Thus green and yellow data points belonging to the same angle $\alpha$
  are not radially connected. Since $\omega_{p}(X=0) = 0$, as explained in the main text, the yellow line coincides with the semi-circle 
  around the green dot corresponding to the zero of $\omega_{p}$. In (b) the angularly averaged value $\overline{\omega}_{p}(X)$ of $\omega_{p}$ for orientations $\alpha \in [0,\pi/2]$ is shown in red.
  For the other scaling variables fixed, $\overline{\omega}_{p}$ represents the critical Casimir potential as function
  of the lateral colloid position $X$, independent of the orientation of the colloid. This average exhibits extrema at the edges of the chemical stripes.
  The sketched Janus particle corresponds to the configuration of minimal energy both with respect to its orientation (see the green line) and its lateral position (see the red line).
  }   
  \label{fig:compass}
\end{figure}

Knowledge of the critical Casimir potential of a cylinder near a chemical step allows one, within DA, to describe more complex geometrical features
of the chemical boundary conditions on the substrate and on the colloid.
Here, we consider a pattern of chemical stripes, which are alternating \textit{p}eriodically along the $x$ direction.
The pattern consists of stripes of width $L_1$ with $(a_1)$ BC neighboring stripes of width
$L_2$ with $(a_2)$ BC such that the periodicity is given by $P=L_1+L_2$.
The coordinate system is chosen such that $x=0$ corresponds to the lateral center of a stripe with $(a_1)$ BC.
Due to the assumed additivity of the forces underlying the DA, within this approximation the critical Casimir potential of a Janus particle as in 
\fref{fig:cyl_substrate}(d) with its center located at a lateral position $x=X$ at a distance $D$ from
such a patterned substrate can be constructed by considering two homogeneous cylinders of half the particle length $L/2$ and summing their scaling functions given in the preceding section. 
We consider this case as an example of certain experimentally relevant geometries, which are difficult to treat 
even within MFT.

The critical Casimir potential $\Phi_p$ acting on a Janus cylinder exhibits the following scaling form:
\small
\begin{equation}
  \label{eq:janus-potential}
	\Phi_p(L_1,P,X,\alpha,D,R,L,T) = k_B T\,\frac{L\,R^{1/2}}{D^{d-3/2}}\,\vartheta_p(\Lambda, \Pi, \Xi, \alpha, \Theta, \Delta, \Len),
\end{equation}
\normalsize
where $\Lambda=L_1/\sqrt{R D}$ and $\Pi=P/\sqrt{R D}$ are, compared with the single chemical step, two additional scaling variables describing the stripe width and the 
periodicity, respectively, and $\vartheta_p$ is the corresponding universal scaling function.

Since the stripe pattern and the surface of the Janus particle are combinations of the $(+)$ and $(-)$ BCs it is convenient to follow \eref{eq:pot-step-split} and introduce the normalized scaling function $\omega_p(\Lambda, \Pi, \Xi, \alpha, \Theta, \Delta, \Len)\in [-1,1]$ such that
\small
\begin{align}
	\label{eq:omega_p-split}
	\vartheta_p(\Lambda,\Pi, \Xi, \alpha, \Theta, \Delta, \Len) =
	&\frac{1}{2} (\vartheta_{(-,-)}(\Theta,\Delta,\Len) + \vartheta_{(+,-)}(\Theta,\Delta,\Len)) \nonumber\\
	&+ \frac{1}{2}\Delta\vartheta(\Theta,\Delta,\Len) \omega_p(\Lambda, \Pi, \Xi, \alpha, \Theta, \Delta, \Len).
\end{align}
\normalsize
Without loss of generality, here we limit the rotation angle $\alpha$ between the chemical steps of the stripes and the axis of the Janus cylinder to the range
$\alpha\in[-\pi/2,\pi/2]$ (see \fref{fig:compass}).
Moreover, we restrict ourselves to the symmetry-breaking BCs $(a_1)=(-)$ and $(a_2)=(+)$ on the substrate as well as on the
two halves of the Janus cylinder. We note that, within DA, $\omega_p$ is independent of this particular choice of BCs.

Within the DA, the scaling functions for the critical Casimir force and the corresponding potential can be constructed via
suitably adding and subtracting scaling functions for the step geometry, analogous to the case of a sphere as described 
in detail in Ref.~\cite{troendle:2010}.
For the sake of brevity, we focus on the normalized scaling function $\omega_p$ [\eref{eq:omega_p-split}] of the critical Casimir potential:\newpage
\small
\begin{align}
\label{eq:omega_p}
\omega_p(\Lambda, &\Pi, \Xi, \alpha, \Theta, \Delta\to0, \Len) = \sum_{n=-\infty}^{\infty}\nonumber\\ \Big\{&\omega_{(+|-,-)}\left(\Xi - \tfrac{\Len}{2} \sin(\alpha) + n\Pi + \tfrac{\Lambda}{2},\alpha,\Theta,\Delta\to 0,\tfrac{\Len}{2}\right) \nonumber\\
-&\omega_{(+|-,-)}\left(\Xi + \tfrac{\Len}{2} \sin(\alpha) + n\Pi + \tfrac{\Lambda}{2},\alpha,\Theta,\Delta\to 0,\tfrac{\Len}{2}\right) \nonumber\\
-&\omega_{(+|-,-)}\left(\Xi - \tfrac{\Len}{2} \sin(\alpha) + n\Pi - \tfrac{\Lambda}{2},\alpha,\Theta,\Delta\to 0,\tfrac{\Len}{2}\right) \nonumber\\
+&\omega_{(+|-,-)}\left(\Xi + \tfrac{\Len}{2} \sin(\alpha) + n\Pi - \tfrac{\Lambda}{2},\alpha,\Theta,\Delta\to 0,\tfrac{\Len}{2}\right)\Big\}.
\end{align}
\normalsize
The sum over $\omega_{(+|-,-)}(\Xi, \alpha,\Theta, \Delta, \Len)$ [\eref{eq:step-pot-da}] 
with appropriate combinations of the first scaling variable takes into account all stripes from $x=-\infty$ to $x=\infty$, and considers four contributions to the 
potential: the half of the Janus particle with $(-)$ BC interacting with stripes of $(+)$ and $(-)$ BCs, and
the other half of the Janus cylinder with $(+)$ BC, which also interacts with stripes of $(+)$ and $(-)$ BCs; 
here we exploit the fact that the potentials for $(+,+)$ and $(-,-)$ BCs are the same.

The resulting scaling function of the potential $\omega_p$ as obtained within the DA ($\Delta\to0$) for $d=3$  
and $\Theta=0$ is shown in \fref{fig:compass} for a cylinder of reduced length $\Len=20$ and for a 
substrate pattern with $L_1=L_2$ and $P=L/4$, so that $\Pi=5$ and $\Lambda=5/2$.
According to our analysis above, for these parameters we expect the DA to provide a good estimate for
the critical Casimir force.

Within the DA, for a Janus particle located opposite to the center of one stripe, i.e., at $X=0$, \linebreak
the scaling function $\omega_p$ of the critical Casimir potential comprises terms \linebreak
$\omega_{(+|-,-)}\left(\pm \tilde{\Xi}_{1,2},\alpha,\Theta,\Delta\to 0,\tfrac{\Len}{2}\right),$
where $\tilde{\Xi}_{1}=\tfrac{\Len}{2} \sin(\alpha) + n\Pi +\tfrac{\Lambda}{2}$ and 
$\tilde{\Xi}_{2}=\tfrac{\Len}{2} \sin(\alpha) + n\Pi -\tfrac{\Lambda}{2}$.\linebreak
Since the scaling function $\omega_{(+|-,-)}$ is an odd function of $\tilde{\Xi}_{1,2}$ and $n\in\mathbb Z$,
there are always two terms in the sum in \eref{eq:omega_p} which cancel each other.
Therefore, the scaling function $\omega_p$ vanishes for $\Xi=0$ and thus the critical Casimir potential does not depend on the orientation 
of the particle as shown by the yellow curve in \fref{fig:compass}(a).
Due to Eqs.~\eqref{eq:cyl-pot-da} and \eqref{eq:omega_p-split}, this corresponds to the potential [\eref{eq:janus-potential}] being the simple average of the potentials of homogeneous cylinders near homogeneous substrates:
\small
\begin{multline}
  \label{eq:average}
  \Phi_p(L_1,P,X=0,\alpha,D,R,L,T) =\\
  \left[\Phi_{(+,+)}(D,R,T) +\Phi_{(+,-)}(D,R,T)\right]/2. 
\end{multline}
\normalsize
\begin{figure}[b!]
  \centering
   \ifTwocolumn
  \includegraphics[width=7.5cm]{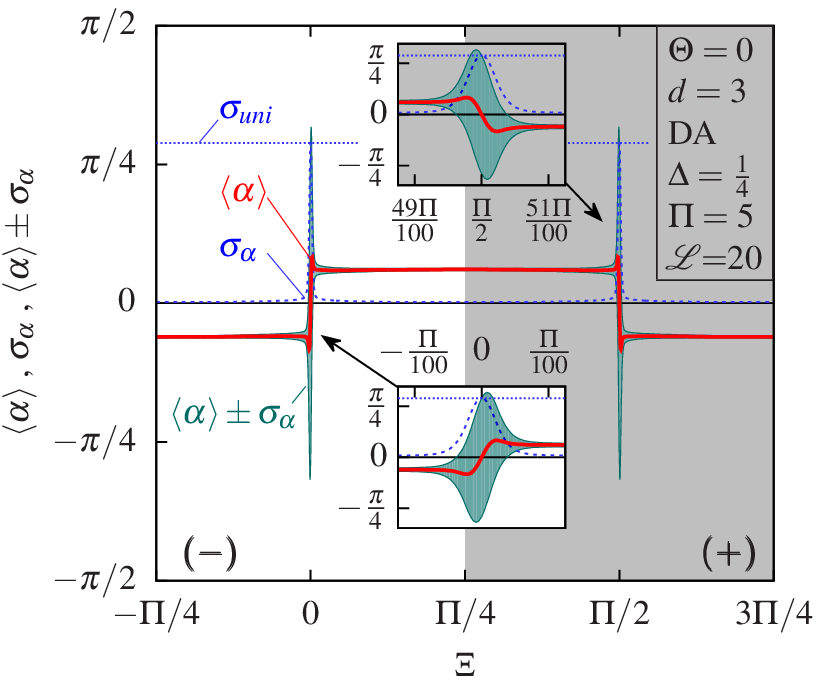}
  \else
  \vspace{-2em}
  \includegraphics[width=11cm]{fig_11}
  \fi
  \clevercaption{%
  Expectation value $\langle\alpha\rangle$ (red curve) and the corresponding standard deviation 
  $\sigma_\alpha$ (blue dashed curve) of the angular probability distribution function 
  $p(\alpha)\propto\exp(-\Phi_p/(k_BT))$ of the same Janus cylinder and for the same parameters as in Fig. \ref{fig:compass}%
  ; there the cylinder is centered at $\Xi=\Pi/4$.
  The dotted blue line denotes the value of the standard deviation $\sigma_{uni}$ of the uniform distribution, up to 
  which, in the present system, $\sigma_\alpha$ grows for specific scaled lateral positions close to $\Xi=m\Pi/2$ with $m\in\mathbb Z$.
  In the insets the green areas are limited by the full green curves $\langle\alpha\rangle\pm\sigma_\alpha$ and visually 
  indicate the areas of the most probable rotation angles $\alpha$ close the lateral positions $\Xi=0$ and $\Xi=\Pi/2$.
  }   
  \label{fig:stripes_moments}
\end{figure}
However, these positions of the colloid center directly above a stripe center are unstable against lateral shifts, 
which can be inferred from, e.g., the yellow square in 
\fref{fig:compass}(b), where we show the value $\overline{\omega}_p(\Xi)$ of the scaling function $\omega_p$
averaged over the tilt angles $\alpha\in[0,\pi/2]$ (red curve).
Therefore $\overline{\omega}_p$ describes the orientationally averaged critical Casimir potential acting on the colloid as a function 
of its lateral position $X$. 
The critical Casimir potential becomes minimal for particle positions at the edges of the stripes, e.g., $X=L_1/2=P/4$, and
with an orientation $\alpha>0$ of the Janus particle such that the overlap of the stripes and of the projected surfaces of the cylinder with equal 
BCs is maximal, as shown in Fig. \ref{fig:compass} (a) by the green curve. 
As a function of $\alpha$ there are also secondary and higher order local minima of the potential, with their number increasing 
for more elongated particles or thinner stripes. 
From our analysis we find, depending on the particle length and the stripe periodicity, $\lceil{L}/(2 P)\rceil$ minima, where $\lceil\ldots\rceil$ indicates the ceiling function.

Equation~\eqref{eq:average} is also obtained in the limit $\Pi=P/\sqrt{R D}\to 0$, so 
that for (infinitely) narrow stripes the angular dependence of the critical Casimir potential disappears.
However, we note that for relatively narrow stripes one has to expect significant deviations from the DA 
due to the increasing interference of the effects of the chemical steps on the order parameter profile across the stripes.
Within MFT in Ref.~\cite{troendle:2010} the range of validity of the DA has been assessed for the case of a spherical colloid next
to a periodically patterned substrate.
Indeed, in Ref.~\cite{parisen:2013} it has been found within a study based both on MC simulations and MFT for the film geometry, that very narrow stripes of alternating $(+)$ and $(-)$ BCs
combine to an effective symmetry-preserving Dirichlet $(o)$ BC.
Nonetheless, for the relatively large value $\Pi=5$ as shown in \fref{fig:compass}, we 
expect the DA to be reliable.

The critical Casimir potential $\Phi_p$ provides the angular probability distribution 
function $p(\alpha)\propto\exp(-\Phi_p/(k_B T))$ characterizing the orientational fluctuations of the cylindrical colloid.
Distinct from the case of a \textit{homogeneous} cylinder near a \textit{single} chemical step, for which we have found a preference for either the parallel 
or the perpendicular orientation, here we observe local minima of the potential (see the green curve in \fref{fig:compass}). 
In order to determine both the preferential particle orientation and the degree of orientational order we calculate the 
moments of the angular probability distribution function
as functions of the reduced lateral position $\Pi$ of the center of the cylinder:
\small
\begin{equation}
\langle\alpha^{n}\rangle = \frac{1}{N} \int\limits_{-\pi/2}^{\pi/2}\upd\alpha\ \alpha^{n}\,e^{-\frac{\Phi_p}{k_B T}},
\end{equation}
\normalsize
where the normalization constant is given by 
$N= \int_{-\pi/2}^{\pi/2}\upd\alpha\ \,\exp\{-\Phi_p/(k_B T)\}$. 
In the following, we employ the usual definitions of the expectation value of $\alpha$ as the first moment 
$\langle\alpha\rangle$ and the standard deviation of the angular distribution 
$\sigma_\alpha = \sqrt{\langle \alpha^2\rangle - \langle \alpha\rangle^2}$. For comparison, the expectation value and the standard 
deviation of the uniform distribution in the interval $[-\pi/2,\pi/2]$ is $\langle\alpha\rangle_{uni}=0$ and $\sigma_{uni} = \pi/(2\sqrt{3})$, respectively. 
These quantities are depicted in \fref{fig:stripes_moments} for the same parameters and the same geometry as in 
\fref{fig:compass}, i.e., for a reduced length $\Len=20$ of the Janus particle and a periodicity $\Pi=5$ of the stripes. 
The expectation value $\langle \alpha \rangle$ is shown in red. It is nearly constant for roughly $48\%$ of 
the first period (i.e., $48\%$ of the range $\Xi\in[-\Pi/4,3\Pi/4]$) and attains a value  
$\langle \alpha \rangle\approx 0.191$ ($\langle \alpha \rangle\approx11^\circ$) at $\Xi=\Pi/4$. 
This nicely agrees with the calculated location $\alpha=\alpha_0=0.192$ of the first minimum as a function of $\alpha$ of the critical Casimir 
potential  at $\Xi=\Pi/4$ and for this particular set of parameters. For $\Xi>\Pi/2$ 
the resulting graph is the mirror image of that for $0<\Xi<\Pi/2$.
Only within a range of $2\%$ of the interval of the period $\Xi \in [-\Pi/4,3\Pi/4]$ 
the expectation value $\langle\alpha\rangle$ deviates noticeably from either $\alpha_0$ or $-\alpha_0$.

In \fref{fig:stripes_moments} the standard deviation $\sigma_\alpha$ is plotted as a blue dashed line. It turns out to be remarkably small for a broad range of 
values of $\Xi$, indicating in that range a very narrow angular distribution around the expectation value. However, for positions close to 
the centers of the stripes, i.e., $\Xi=m\Pi/2$ with $m\in\mathbb Z$, the standard deviation increases, in the present system, up to the value of the uniform distribution $\sigma_{uni}$. 
Consequently, within this $2\%$ range around the centers of the stripes the variation of the expectation value $\langle\alpha\rangle$ does not indicate a change of the orientation but rather a loss of alignment.
As a more intuitive visualization, in \fref{fig:stripes_moments} we also draw $\langle\alpha\rangle\pm\sigma_\alpha$ as full green curves, with the encompassed area shown in light green, indicating the range of the most probable rotation angles $\alpha$. This emphasizes that the Janus particle aligns itself very precisely at a certain angle relative to the pattern, which depends on the stripe width and periodicity, but quite insensitive to the lateral position. Only very close to the center of each stripe the orientation is uniformly distributed. But this is an unstable configuration, as illustrated in \fref{fig:compass}. When the particle is moved laterally over the pattern by external means, its orientation flips between only two preferred alignments $\pm\alpha_0$.

\vspace*{-1em}
\section{Summary and conclusions \label{sec:summary}}
We have studied the critical Casimir interaction between a cylindrical colloid of radius $R$ and length $L$ immersed
in a (near-) critical binary liquid mixture and oriented parallel to a substrate at a surface-to-surface
distance $D$.
In particular, we have focused on several combinations of boundary conditions (BCs) at the colloid and at
the substrate, which are determined by the generic adsorption preferences of surfaces for one of 
the two species of the liquid [\fref{fig:cyl_substrate}].
For chemically homogeneous substrates, a force emerges acting on the colloid along the direction normal
to the substrate.
This force can be attractive or repulsive, depending on the combination of BCs at the colloid and
substrate surfaces.
For chemically inhomogeneous surfaces such as for patterned substrates or Janus particles, 
lateral forces and torques acting on the particles emerge, induced by the critical Casimir effect.
We have calculated the universal scaling functions describing
these multi-directed effective interactions by means of the Derjaguin approximation (DA) in
spatial dimension $d=3$ (based on Monte Carlo results for the universal scaling functions in the film geometry with homogeneous walls) and via mean-field theory (MFT, $d=4$). In the following, we summarize
our main findings.

If the colloid and the substrate exhibit the same homogeneous adsorption preference 
[$(-,-)$ BCs, \fref{fig:cyl_substrate}(a)], the critical Casimir force is attractive. 
The leading universal behavior of the critical Casimir force [\eref{eq:cyl-force-homog}] 
depends on the scaling variables $\Theta$, $\Delta$, and $\Len$ given in \eref{eq:scalingvariables}
and is determined by a universal scaling function shown in \fref{fig:homog_pp} for an
infinitely elongated colloid.
The scaling function derived numerically within MFT supports the validity of the DA 
in the limit of vanishing distance-to-radius ratio $\Delta=D/R\to0$, which
is approached uniformly.
From the comparison with the full MFT result we infer that the actual critical Casimir
force is stronger than the corresponding DA expression for all values of $\Delta$ ranging from
$1/5$ to $6$ and for all temperatures, above and below $T_c$.

In contrast, for opposite adsorption preferences at the substrate and the colloid
[$(+,-)$ BCs, \fref{fig:cyl_substrate}(b)], leading to an effective repulsion, 
the MFT scaling function for the critical Casimir force differs from the DA results even
for small values $\Delta\approx1/5$ [\fref{fig:homog_pm}].
Whereas in the mixed phase of the solvent the DA limit is approached uniformly
and underestimates the force, in the demixed state the DA overestimates the
strength of the critical Casimir force significantly.
This rich behavior of the MFT scaling functions of the critical Casimir force
qualitatively resembles the one reported recently for spherical particles
and obtained from Monte Carlo simulation data \cite{hasenbusch:2013}.

Within our MFT approach we have obtained the full spatial dependence of the 
order parameter profile which describes the solvent.
This has allowed us to interpret these deficiencies of the DA.
For $(+,-)$ BCs and in the demixed state of the liquid an interface
forms, which surrounds the colloidal particle and separates the two phases
of the solvent.
However, according to Figs.~\ref{fig:homog_mod}(a) and \ref{fig:homog_mod}(b) the DA assumes a location and shape of this interface in the region
of closest approach between the colloid and the substrate, which in general
do not agree with the actual ones.
The original DA therefore fails to predict reliably the actual behavior of the critical
Casimir force for this case.
However, based on the knowledge of the interface as determined from the MFT
order parameter profiles, we have presented a modified version of the DA which within MFT overcomes
this deficiency [\eref{eq:Ktilde}].
As shown in \fref{fig:homog_mod}(c), the numerically obtained MFT scaling functions
agree well with this modified DA, which demonstrates that the aforementioned
limitation of the DA is solely due to the shape of the interface formed around 
the colloid.

In Sec.~\ref{sec:finite} we have studied the influence of the finite length of the cylindrical
colloid on the critical Casimir interactions, which for large lengths vanishes $\propto L^{-1}$ [\eref{eq:long}].
Figure~\ref{fig:cyl-finite} shows the dependence of the normal critical Casimir force
on the corresponding scaling variable $\Len=L/\sqrt{RD}$ at the bulk critical point.
From the comparison with the MFT scaling functions we have concluded that already for
$\Len\gtrsim4$ the large-length limit is attained satisfactorily and is rather well
described by the DA, which intrinsically neglects the effects of the cylinder ends
and thus does not depend on $\Len$.
We have concluded that for the parameter ranges $\Delta\lesssim1/3$, $\Len\gtrsim4$,
and in the disordered phase of the liquid the DA reliably predicts the
universal scaling functions for the critical
Casimir force.
We expect this conclusion to carry over to $d=3$ as well.
Thus, in Secs.~\ref{sec:step} and \ref{sec:compass} we focus on this regime.

For a cylindrical colloid exposed to a chemical step where the adsorption
preference of the substrate surface step-like turns into the opposite [\fref{fig:cyl_substrate}(c)],
in Sec.~\ref{sec:step} we have formulated the corresponding scaling behavior
of the critical Casimir forces acting on the colloid both in normal (Sec.~\ref{sec:step}) and in lateral (Appendix \ref{sec:app-lateral}) direction.
In particular the corresponding effective interaction potential shows a
significant dependence on the angle $\alpha$ between the axis of the cylinder 
and the direction of the chemical step, which is shown in \fref{fig:cyltilted_omega} in $d=3$ on
the basis of the DA.

This anisotropy induces a critical Casimir torque [Sec.~\ref{sec:torque}] acting on the cylindrical 
particle [\fref{fig:cyltilted_torque}]. From our analysis we have found that this torque can align the 
colloid parallel or perpendicular to the chemical step, depending on the lateral distance from the step, the 
combination of BCs of the substrate and the colloid, as well as its aspect ratio. In order to analyze the 
degree of orientational order we have investigated the planar nematic order parameter $S$ 
[Figs.~\ref{fig:cyltilted_orderpara} and \ref{fig:cyltilted_orderpara_L2}]. A rod-like particle, with its center located 
above that side of the chemical step sharing its BC ($\Xi>0$), is aligned parallel to the step 
[\fref{fig:cyltilted_orderpara}(a)]. If its center is located above the other side of the chemical step 
with the opposite BC ($\Xi<0$), it prefers the perpendicular orientation in order to increase the overlap of equal BCs
on its surface and on the substrate. Disk-like particles exhibit a different orientational order 
[\fref{fig:cyltilted_orderpara}(c)]. In particular, the preferred orientation of a disk-like colloid 
near a chemical step can be switched reversibly by minute temperature changes [Figs. \ref{fig:cyltilted_orderpara_L2}(a) and \ref{fig:cyltilted_orderpara_L2}(b)], ultimately due to the variation of the ratio of the 
critical Casimir force in the film geometry for $(+,-)$ and $(-,-)$ BCs as a function of temperature.

On the basis of the DA, we have derived these results semi-analytically
in Eqs.~\eqref{eq:pot-step} -- \eqref{eq:pot-delta-theta}, \eqref{eq:step-pot-da}, \eqref{eq:omega-parallel}, and \eqref{eq:cyl_tilted_Ms}.
For the specific cases $\Theta=0$ (corresponding to $T=T_c$) and $\Theta\gg1$ (corresponding
to large $D$ or to large deviations from $T_c$) we have provided analytic expressions
for the corresponding scaling functions in Appendices~\ref{sec:app-crit} and \ref{sec:app-far}.

In Sec.~\ref{sec:compass} we have made use of the general expressions for the critical
Casimir potential derived within DA in order to study the effective interaction 
between a cylindrical Janus particle and a chemically striped substrate.
The effective potential $\Phi_p$ [\eref{eq:janus-potential}] of the colloid exhibits several maxima and minima
depending on the position and the orientation of the particle [\fref{fig:compass}], so that
its preferred axial alignment is rotated relative to the chemical stripes
and shifted laterally with respect to the center of the stripes.
We have characterized the degree of the orientational order using the standard deviation $\sigma_\alpha$ 
of the angular probability distribution function, 
which is surprisingly small except for colloid positions very close to the centers of the chemical stripes. 
A cylindrical Janus particle located at the center of a chemical stripe can rotate de facto freely; but this is an unstable configuration with respect to the lateral position.
The most favorable configuration is achieved when the particle center is positioned 
at the edge of a stripe and aligned as depicted in \fref{fig:compass}. For this particle orientation, the
degree of orientational order is very high  and insensitive to small fluctuations of the particle position.

In summary, the present analysis shows that upon approaching the critical point  
of the solvent, elongated colloidal particles can be reversibly aligned in a designed way 
via minute temperature changes by suitably choosing the geometrical 
parameters of the setup.
Our results provide a means to predict the alignment of
cylindrical colloids near chemically patterned substrates by critical Casimir torques.
Previously, it has been has been demonstrated experimentally that chemically homogeneous 
spherical colloidal particles can be reversibly trapped above a chemically patterned substrate 
via critical Casimir interactions in binary liquid mixtures
\cite{soyka:2008,troendle:2009,troendle:2010,troendle:2011}. 
Using a similar setup, homogeneous cylindrical colloidal particles may be trapped laterally as well as oriented
in a designed way.

Our results indicate that, using the same setup, it is possible to trap Janus cylinders laterally as well as
in an orientation rotated by an angle $\alpha>0$ relative to the substrate symmetry axis, which can be adjusted by 
the geometrical parameters of the substrate pattern in a controlled way.

Such a rich behavior of critical Casimir interactions involving chemically
patterned substrates is also expected to occur for other anisotropic or inhomogeneous colloidal
particles.
This includes particles with ellipsoidal shape \cite{Sacanna:2006,Hu:2008} and dumbbell-shaped particles
\cite{Eisenriegler:2004,Hoffmann:2009,Hoffmann:2010} with homogeneous surfaces as well as
Janus spheres with inhomogeneous surface properties.
The latter type of particles has recently been used for realizing self-propulsion
(see, e.g., Refs.~\cite{Volpe:2011} and \cite{Baraban:2012}).

Critical Casimir forces in colloidal suspensions have been discussed in the literature to 
drive colloidal aggregation phenomena \cite{Beysens:1985,Buzzaccaro:2010,Mohry:2012all}.
Moreover, self assembly of spherical colloidal particles induced by critical Casimir forces has been
observed experimentally in Ref.~\cite{soyka:2008}.
Various colloidal arrangements have been obtained 
depending on the boundary conditions imposed by the underlying chemically patterned substrate \cite{soyka:2008}.
This demonstrates the diverse possibilities of using chemically patterned substrates with a view on 
self-assembly in soft matter.
Our analysis has revealed a rich orientation behavior of anisotropic colloids
near criticality of their solvent
which promises to be useful for designing the alignment of non-spherical colloidal particles.

Our analysis is focused on the leading singular contribution to the forces near $T_c$, which adds
to corresponding nonsingular background interactions due to, inter alia, dispersion and electrostatic forces.
Whereas the latter can be screened by salt, dispersion forces can be weakened by suitable matching of indices
of refraction.
The $(+)$ and $(-)$ BCs can be implemented experimentally by surface treatments only, which leaves the bulk materials 
and thus the leading part of the dispersion forces unchanged.
Therefore, the dispersion forces are to a large extent blind with respect to $(+)$ and $(-)$ BCs, which helps to detect the
critical phenomena discussed here and which crucially live on the contrast between $(+)$ and $(-)$ BCs.
Moreover, the critical phenomena reveal themselves via their singular temperature dependences.
We note that the orientational effects described above should be clearly detectable in experiments due to the strength of the universal contribution of the critical Casimir effect. 
In order to estimate the contrast between attractive and repulsive combinations of BCs, we exemplarily choose the geometry depicted in \fref{fig:cyltilted_orderpara}(a) with $\Delta=1/4$ and $\Len=10$.
For these values, the potential in units of $k_B T$ is given by $\Phi_s/(k_B T)\approx 40\times \vartheta_s$. 
Close to the critical point, the difference of the scaling function $\vartheta_s$ of the potential for $\Xi\to\pm\infty$ is roughly of the order of $1$, leading to a difference of the potential energy of ca. $40\,k_B T$. 
This is in line with previous experiments (see, e.g., Refs.~\cite{hertlein:2008} and \cite{troendle:2011}), and characterizes the universal orientation dependence as being strong compared with the thermal energy and with non-universal background contributions.
%
\appendix
\section{Derjaguin approximation for a cylindrical particle close to a substrate with a chemical step \label{sec:app-da}}
\begin{figure}[b!]
  \centering
  \ifTwocolumn
  \includegraphics[width=7cm]{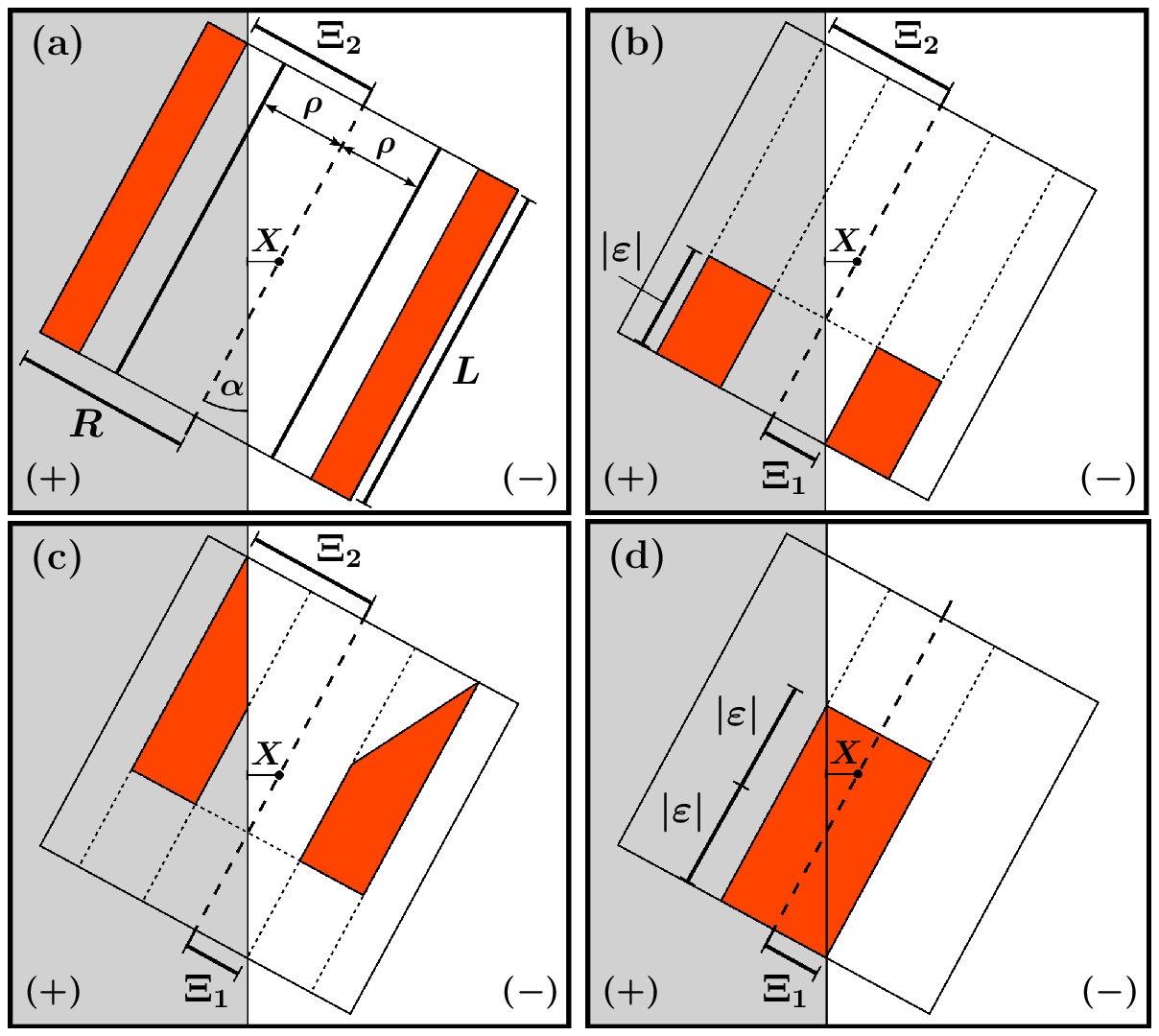}
  \else
  \includegraphics[scale=0.9]{fig_12}
  \fi
    \clevercaption{%
Sketch to illustrate within DA the correspondence of the terms in \eref{eq:step-force-da} to parts of the surface of a cylindrical particle rotated with respect to a chemical step. The rectangular box with thin black lines is the projection of the cylinder surface. The colloid surface is described as a collection of infinitely thin lanes, indicated by the two thick black lines in (a). The variables $\varepsilon$, $\Xi_1$, and $\Xi_2$ in \eref{eq:defs} refer to specific lengths. $\Xi_1$ and $\Xi_2$ designate specific lateral distances from the projected axis of the cylinder where the projection of one of the two lanes forming a pair onto the substrate touches the chemical step with either end. In general, $\Xi_1$ and $\Xi_2$ are different and defined such that $\Xi_2\geq\Xi_1$, with $\Xi_1 = \Xi_2$ only when the colloid center is located exactly above the chemical step, i.e., $X=0$.
In the case of $\varepsilon < 0$, which is depicted here, for a given rotation angle $\alpha$ the colloid is positioned close enough to the chemical step so that the projection of its main axis onto the substrate plane and the chemical step cross each other. If so, $\lvert \varepsilon \rvert$ gives the position of the intersection measured from one end of the cylinder.
The red areas in (a)\ --\ (d) correspond to certain terms as explained in the main text.
    }
    \label{fig:DA_cyl_sketch}
\end{figure}

%
Figure \ref{fig:DA_cyl_sketch} shows the projection of a cylindrical colloid onto a substrate with a chemical step.
Within the DA the surface of the cylinder is sliced into many pairs of infinitely thin lanes, indicated by the two thick black lines in \fref{fig:DA_cyl_sketch}(a). Each pair consists of one lane on the left and one lane on the right of the (dashed) main axis of the cylinder.

In the case of a cylindrical colloid near a chemical step and orientated parallel to it, for some of these pairs both lanes lie such that they are opposite to the same BC on the substrate, whereas for other pairs the two lanes are opposite to different sides of the chemical step, depending on the (scaled) lateral position $\Xi$ of the center of the colloid.

For a cylindrical colloid rotated with respect to the chemical step, the surface lanes may partly overlap both BCs of the substrate. $\Xi_1$ and $\Xi_2$ mark the two lateral distances $\rho$ at which the projected surface lanes touch the chemical step with their ends. The projection of the main axis of the cylinder and the chemical step intersect at a distance $\lvert\varepsilon\rvert$ from the end of the cylinder [see \fref{fig:DA_cyl_sketch}(b)].
If this intersection lies within projected cylinder surface, as depicted in \fref{fig:DA_cyl_sketch}, one has $\varepsilon<0$, otherwise $\varepsilon>0$.
The DA for a cylindrical colloid which is rotated with respect to the chemical step [\eref{eq:step-force-da}] is based on the idea of decomposing the particle surface into areas the contributions of which can be mainly expressed in terms of the known function  $\psi_{(a_<|a_>,b)}^\parallel(\Xi,\Theta)$ [\eref{eq:psi-parallel}].

In order to visualize the contributions from the scaling function $\psi_{(a_<|a_>,b)}^\parallel(\Xi,\Theta)$ [\eref{eq:psi-parallel}], in \fref{fig:DA_cyl_sketch} we color code in red that area corresponding to the integral term, which captures the difference in surface contributions from pairs of lanes with one of these lanes next to a part of the substrate with the same BC as the particle and the other lane next to a part of the substrate with the opposite BC.

For $\rho > \Xi_2$, i.e., the outer part of the projected cylinder surface, shown in red in \fref{fig:DA_cyl_sketch}(a), does not overlap with the chemical step and the contribution from both BCs is given by
\small
\begin{equation}
\psi_{(a_<|a_>,b)}^\parallel(\Xi_2, \Theta) %
= -1 + \int_{1+\Xi_2^2/2}^\infty\upd\beta\,f(\beta)
\label{eq:app_da_a}
\end{equation}
\normalsize
with 
\small
\begin{equation*}
f(\beta)=\sqrt{2}\,(\beta-1)^{-\frac{1}{2}}\,\beta^{-d}\Delta k(\Theta\beta)/\Delta K(\Theta,\Delta\to0,\Len).
\end{equation*}
\normalsize

For $\Xi_1<\rho<\Xi_2$, one part of the surface contributions of the rotated cylinder is equal to the contribution from the partial surface of a shorter cylinder with length $-\varepsilon$ [compare the red area in \fref{fig:DA_cyl_sketch}(b)], corresponding to
\small
\begin{equation}
-\varepsilon \left(\psi_{(a_<|a_>,b)}^\parallel(\Xi_1, \Theta) - \psi_{(a_<|a_>,b)}^\parallel(\Xi_2, \Theta)\right) %
= -\varepsilon \int_{1+\Xi_1^2/2}^{1+\Xi_2^2/2}\upd\beta\,f(\beta) %
\label{eq:app_da_b}
\end{equation}
\normalsize
The other partial contribution for $\Xi_1<\rho<\Xi_2$,
\small
\begin{equation*}
\frac{2\cot|\alpha|}{\Len\Delta K(\Theta,\Delta,\Len)}
  \int_{1+\Xi_1^2/2}^{1+\Xi_2^2/2}\upd\beta\;\beta^{-d}\Delta k(\beta\Theta),
\label{eq:app_da_d}
\end{equation*}
\normalsize
has a different functional form due to the acute angle, as depicted in \fref{fig:DA_cyl_sketch}(c).

The surface contribution for $\rho<\Xi_1$ can be identically rearranged into that of a cylinder with length $2\lvert\epsilon\rvert$ [see the red area in \fref{fig:DA_cyl_sketch}(d)], yielding
\small
\begin{multline}
2\lvert\varepsilon\rvert H(-\varepsilon)\left(\psi_{(a_<|a_>,b)}^\parallel(0, \Theta) - \psi_{(a_<|a_>,b)}^\parallel(\Xi_1, \Theta)\right)\\
= 2\lvert\varepsilon\rvert H(-\varepsilon)\,\varepsilon \int_{1}^{1+\Xi_1^2/2}\upd\beta\,f(\beta) %
\label{eq:app_da_c}
\end{multline}
\normalsize
where $\psi_{(a_<|a_>,b)}^\parallel(0, \Theta)=0$ and $H(x)$ is the Heaviside step function. The sum of these four terms leads to the relative scaling function $\psi_{(a_<|a_>,b)}(\Xi,\alpha,\Theta,\Delta\to0,\Len)$.
The form given in \eref{eq:step-force-da} follows from collecting the coefficients $(\epsilon + 1)$ and $\left(-\varepsilon-2\lvert\varepsilon\rvert H(-\varepsilon)\right)=-\lvert\varepsilon\rvert$ of the the scaling functions $\psi_{(a_<|a_>,b)}^\parallel(\Xi_2, \Theta)$ and $\psi_{(a_<|a_>,b)}^\parallel(\Xi_1, \Theta)$, respectively.

With this decomposition of $\psi_{(a_<|a_>,b)}$, all surface contributions from both substrate BCs, marked by the combined red areas in \fref{fig:DA_cyl_sketch}(a) -- \ref{fig:DA_cyl_sketch}(d), are covered by the integral terms given above. The limiting value for $\Xi\to\pm\infty$ is recovered from $\psi_{(a_<|a_>,b)}^\parallel(\Xi_2\to\pm\infty)\to\mp 1$. The remaining white, trapezoidal area in \fref{fig:DA_cyl_sketch} corresponds to the singular excess interaction of the particle with the right (or left) BC of the chemical step for $\Xi>0$ (for $\Xi<0$).

\section{Scaling functions at $T=T_c$ \label{sec:app-crit}}
%
In this appendix we provide expressions for the scaling functions for the critical Casimir force and potential
acting on a cylindrical colloid next to a chemical step (\fref{fig:cyl_substrate}(c))
as obtained within the DA discussed in Sec.~\ref{sec:step} for the specific case $T=T_c$ and for both $d=3$ and $d=4$ (MFT). 
Right at bulk criticality the scaling function for the film geometry reduces to a universal Casimir amplitude 
[\eref{eq:planar-force}], and, therefore, the integrals presented in Sec.~\ref{sec:step} can be carried out analytically.
We note that the expressions below hold for all combinations of BCs $(a_<)$, $(a_>)$, and $(b)$, independent
of their nature, i.e., symmetry-breaking or symmetry-preserving.
The critical temperature $T=T_c$ corresponds to $\Theta=0$ and $\kab(\Theta=0)=\Dab$, so that the integrals in
Eqs.~\eqref{eq:step-force-da} and \eqref{eq:step-pot-da} can be evaluated for any angle $\alpha$. 
Hence the scaling function $\psi_{(a_<|a_>,b)}$ for the normal critical Casimir force [\eref{eq:step-force-da}] is given by 
(see also  \eref{eq:cyl-force-da})

\small
\begin{multline} 
  \label{eq:psi-0}
  \psi_{(a_<|a_>,b)}(\Xi>0,\alpha,\Theta=0,\Delta\to0,\Len)=
  -|\epsilon|\psi_{(a_<|a_>,b)}^\parallel(\Xi_1,\Theta=0) \\
  +(\epsilon+1)\psi_{(a_<|a_>,b)}^\parallel(\Xi_2,\Theta=0)
  +\frac{2^{6-d} (3 d - 4)}{15\sqrt{2} \pi } \\ \times\left(\left(1+\frac{\Xi_1^2}{2}\right)^{1-d}-\left(1+\frac{\Xi_2^2}{2}\right)^{1-d}\right) \frac{\cot|\alpha|}{\Len}
\end{multline}
\normalsize
with
\small
\begin{multline}
  \label{eq:psi-0-parallel}
  \psi_{(a_<|a_>,b)}^\parallel(\Xi>0,\Theta=0)=
-\frac{2}{\pi}
\Bigg[\arctan\left(\frac{\Xi}{\sqrt{2}}\right) \\
+ \frac{\Xi}{6\sqrt{2}} \frac{\displaystyle\frac{1}{5} (2+16 d)+(-12+5 d) \Xi^2+\frac{3}{2} (-3+d) \Xi^4}{\displaystyle\left(1+\frac{\Xi^2}{2}\right)^{d-1}} \Bigg].
\end{multline}
\normalsize

For the scaling function $\omega_{(a_<|a_>,b)}$ of the critical Casimir potential [\eref{eq:step-pot-da}] we find
\small
\begin{multline}
  \label{eq:omega-0}
  \omega_{(a_<|a_>,b)}(\Xi>0,\alpha,\Theta=0,\Delta\to0,\Len)=
  -|\epsilon|\omega_{(a_<|a_>,b)}^\parallel(\Xi_1,\Theta=0)\\
  +(\epsilon+1)\omega_{(a_<|a_>,b)}^\parallel(\Xi_2,\Theta=0)
  +\frac{2\sqrt{2}(6-d)}{3\pi} \\ \times
  \left[\left(1+\frac{\Xi_1^2}{2}\right)^{2-d}-\left(1+\frac{\Xi_2^2}{2}\right)^{2-d}\right]
  \frac{\cot|\alpha|}{\Len}
\end{multline}
\normalsize
with
\small
\begin{multline}
  \label{eq:omega-0-parallel}
  \omega_{(a_<|a_>,b)}^\parallel(\Xi>0,\Theta=0)\\
  =
  -\frac{2}{\pi } \left[\arctan\left(\frac{\Xi}{\sqrt{2}}\right) + 
  \frac{\displaystyle \left(\tfrac{7}{3} d-6\right)\Xi+ (d-3) \Xi^3}{\displaystyle \sqrt{2}(d-2)\left(1+\frac{\Xi^2}{2}\right)^{d-2}}\right].
\end{multline}
\normalsize
Thus, \eref{eq:omega-0} can be expressed as
\small
\begin{multline}
  \label{eq:omega-0-alt}
  \omega_{(a_<|a_>,b)}(\Xi>0,\alpha,\Theta=0,\Delta\to0,\Len)=
  \frac{2}{\pi}\\
  \Bigg[|\epsilon|\arctan\left(\frac{\Xi_1}{\sqrt{2}}\right)-(\epsilon+1)\arctan\left(\frac{\Xi_2}{\sqrt{2}}\right)\\
  -(d-3)\frac{\sqrt{2}}{3}\left(\left(1+\frac{\Xi_1^2}{2}\right)^{-1}-\left(1+\frac{\Xi_2^2}{2}\right)^{-1}\right)\frac{\cot|\alpha|}{\Len}\Bigg].
\end{multline}
\normalsize
\section{Scaling functions for $\Theta\gg1$ \label{sec:app-far}}
Far from the critical point for (positive) $\Theta\gg1$ \emph{and} for symmetry-breaking BCs 
$(a_<)=(-)$, $(a_>)=(+)$, and $(b)=(-)$, the relationship $\Delta k(\Theta\gg1)=(\mathcal{A}_--\mathcal{A}_+)\Theta^d\exp(-\Theta)$ holds (see \eref{eq:exponential-decay})
and in this limit we find both for $d=3$ and for $d=4$ the same expressions for the scaling functions for the critical
Casimir force, potential, and torque acting on a cylindrical colloid next to a chemical step.

For the scaling functions of the normal critical Casimir force and the corresponding potential we find 
the following same expression (see also Ref.~\cite{troendle:2010}):
\small
\begin{align}
  \label{eq:erf}
  \psi_{(-|+,-)}^\parallel(\Xi>0,\Theta\gg1\Len)&=
  \omega_{(-|+,-)}^\parallel(\Xi>0,\Theta\gg1) \nonumber\\
  &=-\erf\left(|\Xi|\sqrt{\frac{\Theta}{2}}\right).
\end{align}
\normalsize
Thus, from Eqs.~\eqref{eq:step-force-da} and \eqref{eq:psi-parallel} together with the above expression for
 $\Delta k(\Theta\gg1)$ we obtain
\small
\begin{align}
  \label{eq:erf2}%
  \psi_{(-|+,-)}&(\Xi>0,\alpha,\Theta\gg1,\Delta\to0,\Len)\nonumber\\
  =& \omega_{(-|+,-)}(\Xi>0,\alpha,\Theta\gg1,\Delta\to0,\Len)\nonumber\\
  =& |\epsilon|\erf\left(|\Xi_1|\sqrt{\tfrac{\Theta}{2}}\right)
  -(\epsilon+1)\erf\left(|\Xi_2|\sqrt{\tfrac{\Theta}{2}}\right)\nonumber\\
  &+\sqrt{\frac{2}{\pi\Theta}}\frac{\cot|\alpha|}{\Len}\left(\exp\left(-\tfrac{\Xi_2^2}{2}\Theta\right)-\exp\left(-\tfrac{\Xi_1^2}{2}\Theta\right)\right).
\end{align}
\normalsize

\section{Critical Casimir torque \label{sec:app-torque}}
In the following, we determine the scaling function $M_s$ for the critical Casimir 
torque acting on a cylindrical colloid next to a chemical step (\fref{fig:cyl_substrate}(c))
as obtained within the DA discussed in Sec.~\ref{sec:step}.
In this limit, the corresponding derivative of $\omega_{(a_<|a_>,b)}$, as expressed in
\eref{eq:msdef}, is given by
\begin{subequations}
  \small
  \begin{equation}
    \frac{d}{d\alpha}\, \omega_{(a_<|a_>,b)}=\\%
    \left[
    \frac{\partial}{\partial \alpha} 
    + \frac{d\Xi_1}{d\alpha} \frac{\partial}{\partial \Xi_1} 
    + \frac{d\Xi_2}{d\alpha} \frac{\partial}{\partial \Xi_2} 
    + \frac{d\epsilon}{d\alpha} \frac{\partial}{\partial \epsilon} 
    \right]
    \omega_{(a_<|a_>,b)},%
  \end{equation}
  \normalsize
  where for $0\leq\alpha\leq \pi/2$
  \small
  \begin{align}
    \frac{d\Xi_1}{d\alpha} &= \sgn(\epsilon)\frac{\lvert\Xi\rvert \sin\alpha-\Len / 2}{\cos^2 \alpha},\nonumber\\ 
    \frac{d\Xi_2}{d\alpha} &= \frac{\lvert\Xi\rvert \sin\alpha+\Len / 2}{\cos^2 \alpha}, \nonumber\\
    \frac{d\epsilon}{d\alpha} &= - \frac{\lvert\Xi\rvert \cot\alpha}{\Len \sin\alpha}, 
  \end{align}
  \normalsize
and
\small
  \begin{equation}
    \frac{\partial\omega_{(a_<|a_>,b)}}{\partial \alpha}  = - \frac{2}{\Len\Delta\vartheta(\Theta,\Delta\to0,\Len)\sin^2 \alpha}\, I_\omega(\Xi_1, \Xi_2,\Theta),
  \end{equation}
  \begin{equation}
    \frac{\partial\omega_{(a_<|a_>,b)}}{\partial \epsilon}  = \omega^\parallel_{(a_<|a_>,b)}(\Xi_2,\Theta)-\sgn(\epsilon)\omega^\parallel_{(a_<|a_>,b)}(\Xi_1,\Theta),
  \end{equation}
  \begin{equation}
    \frac{\partial\omega_{(a_<|a_>,b)}}{\partial \Xi_1}  = \frac{\partial\omega_{(a_<|a_>,b)}}{\partial \Xi_2} = 0. \label{eq:omega-derivs-xi12}
  \end{equation}
\normalsize
  \label{eq:final1}
\end{subequations}
Note that \eref{eq:omega-derivs-xi12} is not obvious a priori, but carefully differentiating $\omega_{(a_<|a_>,b)}$ in \eref{eq:step-pot-da} with respect to $\Xi_1$ and $\Xi_2$ reveals that all emerging terms cancel each other.

Thus, from Eqs.~\eqref{eq:msdef} and \eqref{eq:final1} we find the following expression for the scaling function of the critical Casimir torque in the limit $\Delta\to 0$:
\small
\begin{align}
  M_s(\Xi\gtrless 0,\alpha, \Theta,\Delta\to0,\Len) =
&\mp \frac{1}{\sin^2\alpha}\Bigg\{ \frac{1}{\Len}\,I_\omega(\Xi_1, \Xi_2,\Theta) \nonumber\\
  & + \frac{\Delta\vartheta(\Theta,\Delta\to0,\Len)}{2}
  \frac{\lvert\Xi\rvert}{\Len}\cos\alpha \nonumber\\
  &\times
   \Bigg[ \omega^\parallel_{(a_<|a_>,b)}(\Xi_2,\Theta) \nonumber\\
          & -\sgn(\epsilon)\omega^\parallel_{(a_<|a_>,b)}(\Xi_1,\Theta)
          \Bigg]\Bigg\}.
\label{eq:cyl_tilted_Ms-app}
\end{align}
\normalsize

At the critical point $T=T_c$, the scaling function $\omega_{(a_<|a_>,b)}$ of the critical Casimir
potential is given by Eqs.~\eqref{eq:omega-0} - \eqref{eq:omega-0-alt} in Appendix \ref{sec:app-crit}. 
Hence at $T_c$, the scaling function $M_s$ for the critical Casimir torque [\eref{eq:cyl_tilted_Ms-app}] reduces to
\small
\begin{align}
\label{eq:cyl_tilted_Ms_bulk}
  M_s(\Xi\gtrless 0 &,\alpha, \Theta=0,\Delta\to0,\Len) =
  \mp\frac{\Delta\vartheta(\Theta=0,\Delta\to0,\Len)}{2\Len \sin^2\alpha} \nonumber\\ 
  &\times\Bigg \{
    {\lvert\Xi\rvert \cos\alpha} 
    \Big[ \omega^\parallel_{(a_<|a_>,b)}(\Xi_2,\Theta=0) \nonumber\\
          &-\operatorname{sgn}(\epsilon)\omega^\parallel_{(a_<|a_>,b)}(\Xi_1,\Theta=0)
    \Big] \nonumber\\
    &+ \frac{4\sqrt{2}}{\pi(d-1)}\left(\left(1+\frac{\Xi_1^2}{2}\right)^{2-d}-\left(1+\frac{\Xi_2^2}{2}\right)^{2-d}\right)\Bigg\}
\end{align}
\normalsize
with $\omega^\parallel_{(a_<|a_>,b)}(\Xi_{1,2},\Theta=0)$ given by \eref{eq:omega-0-parallel}.

Far from the critical point for (positive) $\Theta\gg1$ \emph{and} for symmetry-breaking BCs 
$(a_<)=(-)$, $(a_>)=(+)$, and $(b)=(-)$, we obtain with \eref{eq:erf}
\small
\begin{align}
M_s(\Xi\gtrless 0 &,\alpha, \Theta\gg 1,\Delta\to0,\Len) =
 \mp\frac{\Delta\vartheta(\Theta\gg 1,\Delta\to0,\Len)}{2\Len \sin^2\alpha} \nonumber\\
& \times \Bigg \{-\lvert\Xi\rvert \cos\alpha\left[\erf\left(\Xi_2\sqrt{\tfrac{\Theta}{2}}\right)-\sgn(\epsilon)\erf\left(\Xi_1\sqrt{\tfrac{\Theta}{2}}\right)\right] \nonumber\\
& + \sqrt{\frac{2}{\pi\Theta}}\left[\exp\left(-\tfrac{\Xi_1^2}{2}\Theta\right)-\exp\left(-\tfrac{\Xi_2^2}{2}\Theta\right)\right]\Bigg\}.
\label{eq:cyl_tilted_Ms_far}
\end{align}
\normalsize
\vspace*{-2em}
\section{Lateral critical Casimir force \label{sec:app-lateral}}
For completeness, in this appendix we present the
expression for the \textit{l}ateral critical Casimir force acting on 
a cylindrical colloid in the 
proximity of a chemical \textit{s}tep [\fref{fig:cyl_substrate}(c)].
The dependence of the critical Casimir potential on the particle position $X$
relative to the chemical step induces a force acting on the particle along the $x$ direction parallel to
the substrate. This lateral critical Casimir force
$F_s^{\lat}$ for a cylindrical particle follows from the
potential energy $\Phi_s$ according to $F_s^{\lat}\equiv-\frac{d}{d X}\Phi_s$.
It can be written in the  scaling form (compare \eref{eq:force-step})
\small
\begin{equation}
  \frac{F_s^{\lat}(X,\alpha,D,R,L,T)}{k_B T}= \frac{L R^{1/2}}{D^{d-1/2}} \left(\frac{D}{R}\right)^{1/2} K_s^{\lat}(\Xi,\alpha,\Theta,\Delta,\mathcal L),
\end{equation}
\normalsize
with the scaling functions $K_s^{\lat}$ (omitting the index $(a_<|a_>,b)$) of the lateral
critical Casimir force, which follows from
the scaling functions of the potential in \eref{eq:pot-step} and \eqref{eq:pot-step-split}:
\small
\begin{equation}
  \label{eq:lateral-K}
  K_s^{\lat}(\Xi,\alpha,\Theta,\Delta,\mathcal L)\\ 
  = -\frac{1}{2}\Delta\vartheta(\Theta,\Delta,\mathcal L)
  \frac{d}{d\Xi}\omega_{(a_<\vert a_>,b)}(\Xi,\alpha, \Theta, \Delta, \mathcal L).
\end{equation}
\normalsize

Within the DA, i.e., for $\Delta\to 0$, in \eref{eq:step-pot-da} the expression for $\omega_{(a_<|a_>,b)}(\Xi,\alpha, \Theta, \Delta\to0,\mathcal L)$ has been provided in an implicit form
using the variables $\epsilon$, $\Xi_1$, and $\Xi_2$, which depend on $\Xi$, $\alpha$, and $\mathcal L$ 
[\eref{eq:defs}].
Thus, in \eref{eq:lateral-K} the total derivative is
\begin{subequations}
\small
\begin{equation}
  \frac{d}{d\Xi}\omega_{(a_<|a_>,b)}= \\
  \left[\frac{\partial}{\partial \Xi} + \frac{d\Xi_1}{d\Xi} \frac{\partial}{\partial \Xi_1} 
  + \frac{d\Xi_2}{d\Xi} \frac{\partial}{\partial \Xi_2} 
  + \frac{d\epsilon}{d\Xi} \frac{\partial}{\partial \epsilon} \right]\omega_{(a_<\vert a_>,b)},
\end{equation}
\normalsize
where
\small
\begin{equation}
\frac{d\Xi_1}{d\Xi} = \frac{\sgn(\epsilon)\sgn(\Xi)}{\cos \alpha}, \; \frac{d\Xi_2}{d\Xi} = \frac{1}{\cos \alpha}, \; \text{and } \frac{d\epsilon}{d\Xi} = \frac{\sgn(\Xi)}{\mathcal L \sin \alpha}.
\end{equation}
\normalsize
By using the relationships
\small
\begin{equation}
\frac{\partial\omega_{(a_<|a_>,b)}}{\partial \Xi}  = 0
\end{equation}
\normalsize
and
\small
\begin{equation} 
\frac{\partial\omega_{(a_<|a_>,b)}}{\partial \Xi_1} =
\frac{\partial\omega_{(a_<|a_>,b)}}{\partial \Xi_2} = 0,
\end{equation}
\normalsize
[see \eref{eq:final1}], as well as
\small
\begin{multline}
\frac{\partial\omega_{(a_<|a_>,b)}}{\partial \epsilon}  = \sgn(\Xi)\\
\times\left(\omega^\parallel_{(a_<|a_>,b)}(\Xi_2,\Theta)-\sgn(\epsilon)\omega^\parallel_{(a_<|a_>,b)}(\Xi_1,\Theta)\right),
\label{eqn:cyl_tilted_omega_derivs}
\end{multline}
\normalsize
\end{subequations}
the scaling function of the lateral critical Casimir force $K_s^{\lat}$ within the DA, which is a symmetric function of $\Xi$, can finally be expressed as
\small
\begin{multline}
\label{eq:cyl_tilted_Ks_func}
K_s^{\lat}(\Xi,\alpha,\Theta,\Delta\to0,\mathcal L) =
\frac{\Delta\vartheta(\Theta,\Delta\to0,\mathcal L)}{2 \mathcal L \sin \alpha} \\
\times
\Bigg[\omega^\parallel_{(a_<|a_>,b)}(\Xi_2,\Theta) 
-\sgn(\epsilon)\omega^\parallel_{(a_<|a_>,b)}(\Xi_1,\Theta)\Bigg].
\end{multline}
\normalsize
\section*{Acknowledgements}
The authrors thank Andrea Gambassi for helpful discussions.

%
%
%
%

\end{document}